# Structural phase transitions and photoluminescence mechanism in a layer of 3D hybrid perovskite nanocrystals

Yuri D. Glinka,[1,2,a)] Rui Cai,[1] Xian Gao,[1] Dan Wu,[1] Rui Chen,[1,b)] and Xiao Wei Sun[1,3,c)]

[1]Guangdong University Key Lab for Advanced Quantum Dot Displays and Lighting, Shenzhen Key Laboratory for Advanced Quantum Dot Displays and Lighting, Department of Electrical & Electronic Engineering, Southern University of Science and Technology, Shenzhen 518055, China
[2]Institute of Physics, National Academy of Sciences of Ukraine, Kiev, 03028, Ukraine.
[3]Shenzhen Planck Innovation Technologies Pte Ltd, Ganli 6th Road, Longgang, Shenzhen 518112, China

[a)]yuri@sustc.edu.cn, yuridglinka@yahoo.com
[b)]chenr@sustc.edu.cn
[c)]sunxw@sustc.edu.cn

Although the structural phase transitions in single-crystal hybrid methyl-ammonium (MA) lead halide perovskites ($MAPbX_3$, X = Cl, Br, I) are common phenomena, they have never been observed in the corresponding nanocrystals. Here we demonstrate that two-photon-excited photoluminescence (PL) spectroscopy is capable of monitoring the structural phase transitions in $MAPbX_3$ nanocrystals compared to conventional one-photon-excited PL spectroscopy because nonlinear susceptibilities govern the light absorption rates. Using this technique, we provide experimental evidence that the orthorhombic-to-tetragonal structural phase transition in a single layer of 20-nm-sized 3D $MAPbBr_3$ nanocrystals is spread out within the ~70 - 140 K range. This structural phase instability range arises because, unlike in single-crystal $MAPbX_3$, free rotations of MA ions in the corresponding nanocrystals are no longer restricted by a long-range MA dipole order. The resulting configurational entropy loss and the corresponding liquid-like motion of MA cations can be even enhanced by the interfacial electric field arising due to charge separation at the $MAPbBr_3$/ZnO heterointerface, extending the orthorhombic-to-tetragonal structural phase instability range from 70 to 230 K. We conclude that the weak sensitivity of conventional one-photon-excited PL spectroscopy to the structural phase transitions in 3D $MAPbX_3$ nanocrystals results from the structural phase instability providing negligible distortions of $PbX_6$ octahedra responsible for the band-edge electronic transitions. In contrast, the intensity of two-photon-excited PL and electric-field-induced one-photon-excited PL still remains sensitive enough to weak structural distortions due to the higher rank tensor nature of nonlinear susceptibilities involved. We also show that room-temperature PL originates from the radiative recombination of the optical-phonon vibrationally excited polaronic quasiparticles with energies might exceed the ground-state Fröhlich polaron and Rashba energies due to optical-phonon bottleneck. Because of small masses and large radii of these vibrationally excited polaronic quasiparticles, their high mobility and long-range diffusion become possible.

## I. INTRODUCTION

Hybrid methyl-ammonium (MA) lead halide perovskites ($MAPbX_3$, X = Cl, Br, I) represent a class of materials offering an illustrative platform for studying the relaxation dynamics of photoexcited carriers and their transport phenomena in novel highly efficient solar cells for solar energy harvesting technology.[1-11] One of the most important specific features of these hybrid materials is that their crystalline structure can be viewed as two alternating sublattices. Specifically, the inorganic sublattice is composed by corner-sharing $PbX_6$ octahedra which are responsible for forming the valence band (VB) maximum and conduction band (CB) minimum of these materials.[7,8] Consequently, the initial relaxation of photoexcited carriers, their recombination and transport phenomena all occur in the inorganic sublattice. Alternatively, the organic MA sublattice acts as a medium modifying the electrostatic potential of the inorganic sublattice, thus contributing less significantly to the charge screening and localization effects, nevertheless, providing an ultralow thermal conductivity being caused by long-range MA dipole order.[12] The structural peculiarities of these materials allow for the three structural phase transitions occurring in the temperature range of $T$ ~140 – 240 K, which usually appear in single-crystal $MAPbX_3$ and its polycrystalline thin film,[13-25] whereas they have never been observed in $MAPbX_3$ nanocrystals.[21]

Because arrays of colloidal nanocrystals are known to be promising alternatives to the single-crystal semiconductor based electronics, optoelectronics, and solar energy harvesting applications[26] and because the properties of single-crystal $MAPbX_3$ differ significantly for different structural phases,[13-25] we comprehensively explored the structural phase transitions in a fully encapsulated single layer of 20-nm-sized 3D $MAPbBr_3$ nanocrystals using one-photon-excited and two-photon-excited PL spectroscopy. The effect of the technologically important $MAPbBr_3$/ZnO heterointerface on this phenomenon has also been studied here. We show that two-photon-excited PL spectroscopy and electric-field-induced one-photon-excited PL spectroscopy are capable of more precisely monitoring the structural phase transitions in 3D $MAPbBr_3$ nanocrystals compared to conventional one-photon-excited PL spectroscopy since nonlinear susceptibilities govern the light absorption rates. Consequently, one can recognize that the orthorhombic-to-tetragonal structural phase transition in 3D $MAPbBr_3$ nanocrystals, unlike in single-crystal $MAPbX_3$, is spread out over the broad temperature range of $T$ ~70 - 140 K. This extension of the structural phase transition defines the structural phase instability range, within which the local field fluctuations arising due to free rotations of MA ions are no longer restricted by long-range polar order.[27,28] The resulting configurational entropy loss and the corresponding liquid-like motion of MA cations[1] can be even enhanced by the interfacial electric field when charge separation at the $MAPbBr_3$/ZnO heterointerface occurs, extending the orthorhombic-to-tetragonal structural phase instability range from $T$ ~70 to 230 K. The latter effect is found to be dependent on the ZnO layer thickness and the photoexcited carrier density. Finally, we conclude that a stepwise shift of the PL band with temperature observed for single-crystal $MAPbX_3$ and assigned to structural phase transitions does not appear anymore in 3D $MAPbX_3$ nanocrystals because of negligible distortions of $PbX_6$ octahedra under the structural phase instability regime. On the contrary, the nearly monotonic blue shift of PL band with increasing temperature in a fully encapsulated single layer of 20-nm-sized 3D $MAPbBr_3$ nanocrystals seems to result rather from the heating effect under TO/LO phonon bottleneck[2-4] than that being induced by the progressive $PbX_6$ octahedra distortions. Consequently, room-temperature PL is expected to originate from the radiative recombination of the optical-phonon vibrationally excited polaronic quasiparticles with energies might exceed the ground-state Fröhlich polaron and Rashba energies due to optical-phonon bottleneck. Because of small masses and large radii of these vibrationally excited polaronic quasiparticles, their high mobility and long-range diffusion become possible.

## II. EXPERIMENTAL

### A. Sample fabrication

The size-controlled $CH_3NH_3PbBr_3$ nanocrystals were synthesized by a ligand-assisted reprecipitation (LARP) technique.[29] The $CH_3NH_3Br$ precursor was synthesized by adding hydrobromic acid (HBr, 48 wt.% in $H_2O$, 99.99%; Sigma-Aldrich) drop wise to a stirred solution of methylamine ($CH_5N$, 30~33 wt.% in ethanol) at 0 ºC followed by stirring for 1 h. Upon drying at 100 °C in air, white $CH_3NH_3Br$ powder in quantitative yield was formed. After being washed with diethyl ether (Shanghai Lingfeng Chemical Reagents) and re-crystallized with ethanol, $CH_3NH_3Br$ powder was dried for 24 h in a vacuum furnace and along with other precursors was added into the N,N-dimethylformamide (DMF, $C_3H_7NO$, anhydrous, 99.8%). Specifically, $CH_3NH_3Br$ and lead (II) bromide ($PbBr_2$, powder, 98%) were dissolved in 100 µL DMF forming a mixture with a concentration of 0.1 mM and then 200 µL oleic acid ($C_{18}H_{34}O_2$; Aladdin) and 20 µL oleylamine ($C_{18}H_{37}N$, 80~90%; Aladdin) were added into this mixture. The oleic acid/oleylamine ligand ratio was selected for tailoring the size of the nanocrystals. The 100 µL mixture of various precursors was injected afterwards into 3 mL chloroform (Shanghai Lingfeng Chemical Reagents) as an antisolvent. A yellow-greenish colloidal solution was acquired afterwards. For further purification, 1.5 mL toluene/acetonitrile ($CH_3CN$, anhydrous, 99.8%) mixture with a volume ratio 1:1 was added into the solution and the sediment was dispersed in hexane after centrifuging at 9000 rpm for 2 min.

To prepare the fully encapsulated layer of 3D $CH_3NH_3PbBr_3$ nanocrystals, the sapphire plates (10×10×0.3 mm; Jiangsu Hanchen New Materials) were cleaned by successively soaking them in an ultrasonic bath with deionized water, acetone, and isopropanol for 10 min each and dried with nitrogen. The sapphire substrates were transferred afterwards into the atomic layer deposition (ALD) system (PICOSUN™ R-200) to grow a ZnO film. Diethyl zinc (DEZn, $Zn(C_2H_5)_2$) and $H_2O$ were used as precursors. High purity nitrogen with dew point below -40 ºC was used as a purging and carrier gas. The reactor chamber pressure was set as 1000 Pa during the growth. When the growth temperature of 200 °C was reached, DEZn was introduced to the reactor chamber with a flow rate of 150 sccm followed by purging with nitrogen to remove the residues and byproducts. The precursor of $H_2O$ with a flow rate of 200 sccm was introduced afterwards into the reactor chamber to start with the ZnO layer growth. The number of ALD growth cycles was selected to grow a ZnO layer with thicknesses of 30 and 100 nm, which were also verified by other methods.

Closely packed and uniformly distributed $CH_3NH_3PbBr_3$ nanocrystals were spin-coated afterwards by optimizing the spin speed to 1500 rpm to either the clean sapphire (Sa) plate or to those initially ALD-coated with a ZnO layer of 30 nm and 100

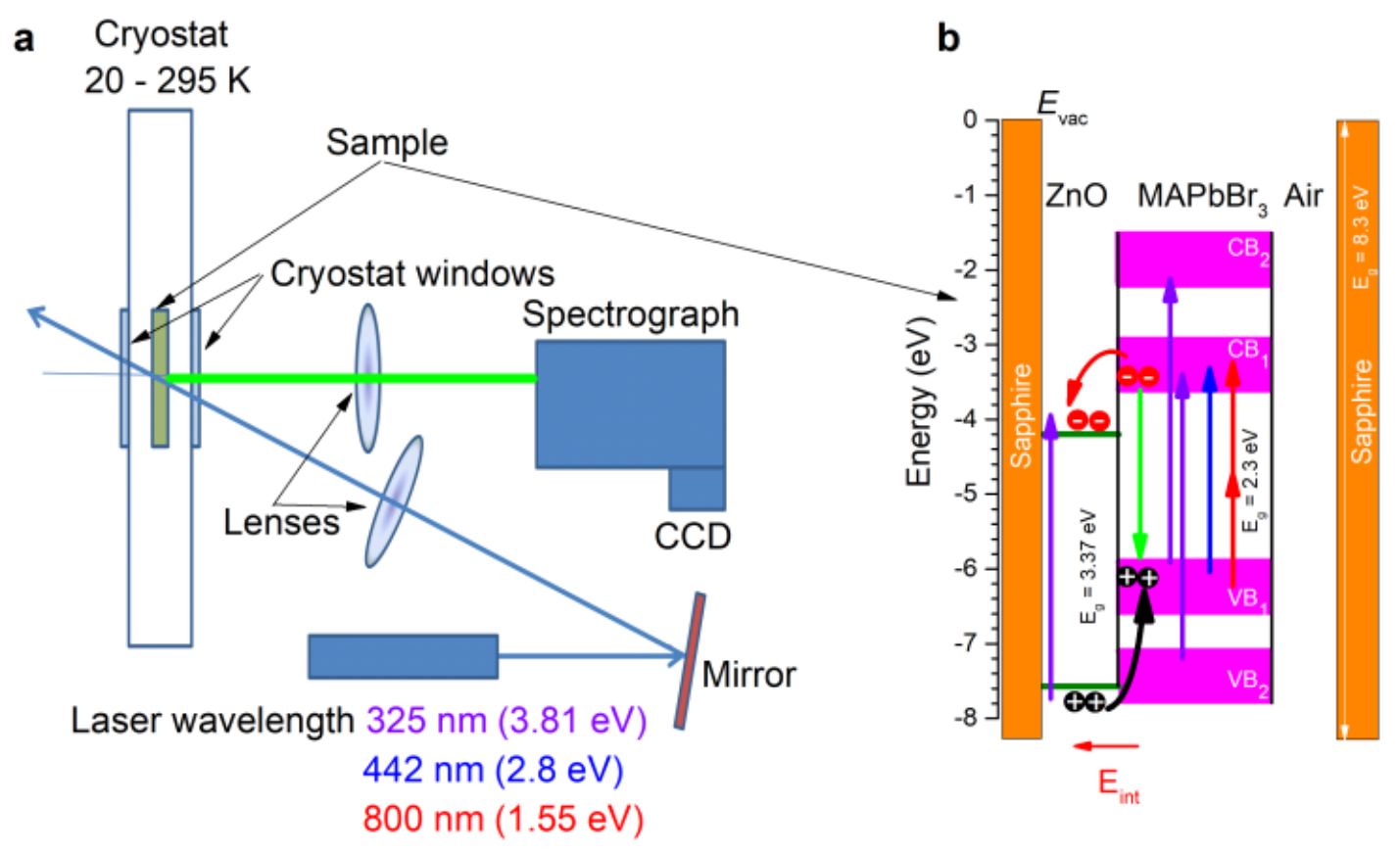

**FIG. 1** (a) A sketch of the experimental setup. (b) A band alignment of a fully encapsulated layer of $MAPbBr_3$ nanocrystals spin-coated to the ALD-grown ZnO layer. The charge separation process at the $MAPbBr_3$/ZnO heterointerface and the resulting interfacial electric field ($E_{int}$) are shown. Laser wavelengths indicated in (a) and the corresponding electronic transitions indicated in (b) are shown by the same colors.

nm thickness. The resulting structure was covered by another sapphire plate, leaving the air gap above the nanocrystal array of ~1 μm and gluing sapphire plates on sides by UV adhesive. The obtained samples will be referred further below as $MAPbBr_3$/Sa, $MAPbBr_3$/ZnO(30nm) and $MAPbBr_3$/ZnO(100nm), respectively.

**B. Scanning electron microscopy (SEM) imaging.**
The cross-sectional SEMs images of the sandwiched samples were acquired using a ZEISS Gemini 300 field emission scanning electron microscope in a secondary electron mode after cleaving the samples with a diamond scriber.

**C. Transmission electron microscopy (TEM) imaging**
The crystallinity of the synthesized $CH_3NH_3PbBr_3$ nanocrystals was confirmed by TEM imaging (Tecnai F30 field-emission TEM) operated at 300 kV and at room temperature.

**D. X-ray diffraction (XRD) characterization**
The XRD patterns of the synthesized $CH_3NH_3PbBr_3$ nanocrystals were measured using a Rigaku SmartLab X-ray diffractometer, equipped with a Cu KR radiation source (wavelength at 1.542 Å). The samples were scanned from $10^o < 2\theta < 60^o$ at an increment of $10^o$/min.

**E. Conventional ultraviolet–visible absorption and PL characterization**
The conventional absorption spectra were measured at room temperature using the Beijing Spectrum Analysis 1901 Series spectrometer. To study PL spectra, the Ocean Optics QE 65 Pro spectrometer equipped with a 365 nm excitation source was used with a spectral resolution of ~1.0 nm.

**F. Temperature-dependent PL measurements**
To study temperature-dependent PL spectra, the commercially available temperature controller (Lakeshore 336) with a temperature range of 20 - 295 K was used. The PL spectra were measured using the Andor Shamrock SR750 spectrograph equipped with a CCD detector. Figure 1a shows a sketch of the experimental setup for measuring the temperature dependent PL from a layer of 3D $MAPbBr_3$ NCs. Three laser wavelengths were used to excite PL: 325, 442, and 800 nm, which were emitted from the CW He-Cd laser (325 and 442 nm) and the femtosecond laser (Astrella-Tunable-V-F-1K) with the pulse width of 100 fs and a repetition rate of 1.0 KHz (800 nm). The laser beam incidence angle was 30°. The sample holder was designed to allow the excitation laser light to pass through the sample and the cryostat windows, being blocked afterwards outside the cryostat. Such an arrangement eliminates any additional spectroscopic features associated with the scattering and reabsorption of laser light from the sample holder to appear in PL spectra. The laser spot diameter was ~250 μm for all the three excitation wavelengths. The excitation power was changed by a variable neutral density filter (Thorlabs). The averaged laser power varied from 0.03 to 20 mW for the CW He-Cd laser and from 2.0 to 30 mW for the pulsed laser. For the experimental conditions applied, 1.0 mW average laser power corresponds to the laser light intensity (power density) of 2.04 W/$cm^2$ for the CW laser and 20.5 GW/$cm^2$ for the pulsed laser. Taking into account the measured one-photon absorption coefficient (~$4.0 \times 10^4$ $cm^{-1}$ and ~$1.9 \times 10^4$ $cm^{-1}$ for 325 and 442 nm laser light, respectively) and the two-photon absorption coefficient of ~8.6 cm/$10^9$ W,[30] the reflectance coefficient of 0.37 and estimating afterwards the power density absorbed within the $CH_3NH_3PbBr_3$ nanocrystal array,[31] the corresponding photoexcited carrier densities were calculated to be ~$7.5 \times 10^{17}$ $cm^{-3}$ ( ~$7.5 \times 10^{-4}$ $nm^{-3}$), ~$4.9 \times 10^{17}$ $cm^{-3}$ ( ~$4.9 \times 10^{-4}$ $nm^{-3}$), and ~$1.7 \times 10^{18}$ $cm^{-3}$ (~$1.7 \times 10^{-3}$ $nm^{-3}$) for 325, 442, and 800 nm laser light, respectively. Assuming the nanocrystal cubic shape of the same edge length of ~20 nm (the corresponding volume is ~$8 \times 10^3$ $nm^3$), the average electron-hole pair occupancy per nanocrystal can be estimated as ~6.0, ~3.9, and ~13.6, respectively. However, multiple excitons photoexcited in a nanocrystal is believed to be non-interactive since their small effective exciton Bohr radius in $MAPbX_3$ perovskite materials (2.0 – 4.0 nm),

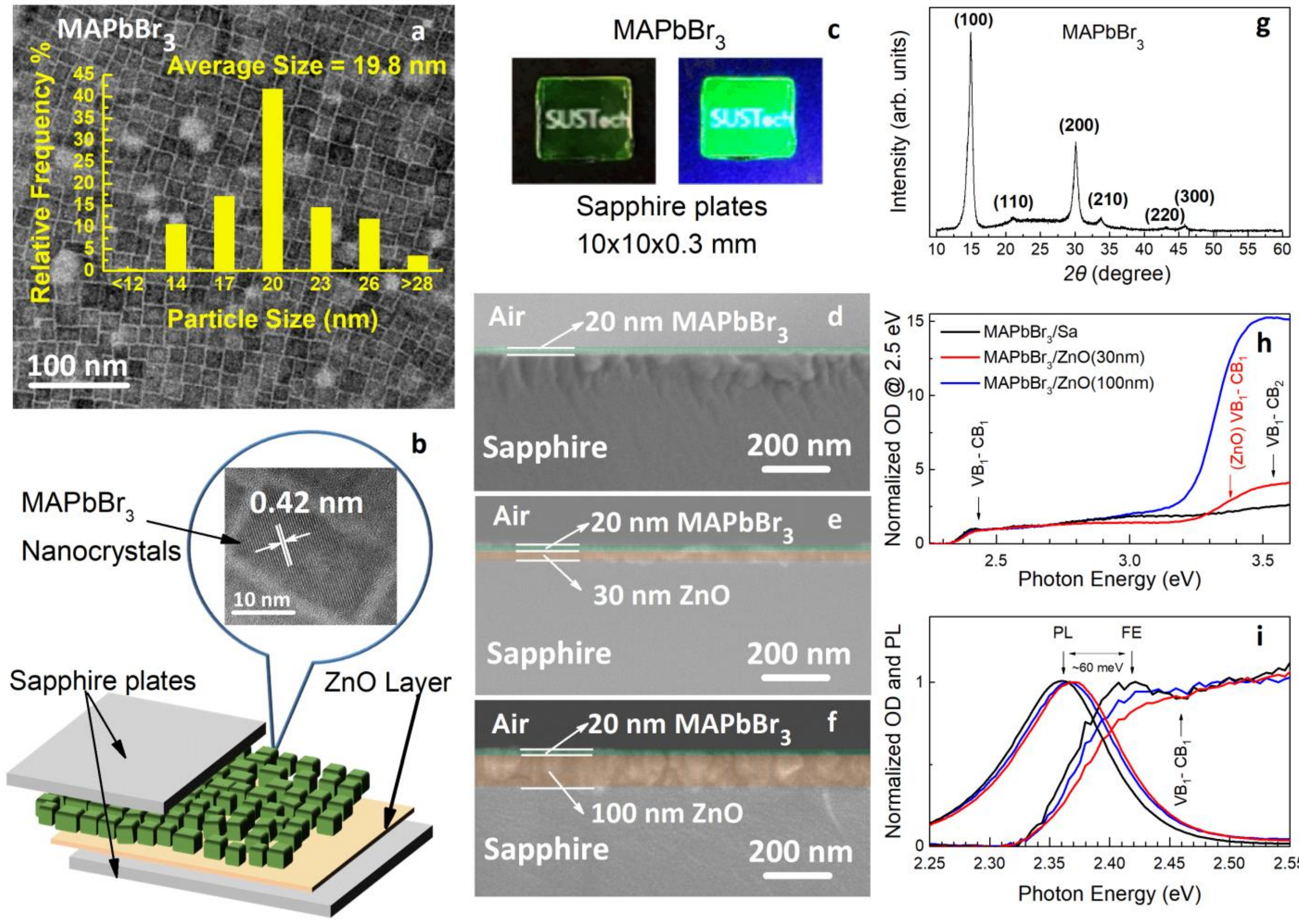

**FIG. 2** (a) The TEM image of 3D $MAPbBr_3$ nanocrystals. The corresponding histogram demonstrates the nanocrystal size distribution. (b) A schematic presentation of the encapsulated layer of 3D $MAPbBr_3$ nanocrystal and the high-resolution TEM image of an individual $MAPbBr_3$ nanocrystal. (c) A real image of a layer of 3D $MAPbBr_3$ nanocrystals illuminated by daylight (left) and UV light (right). The white color "SUSTech" label placed behind the sample demonstrates its transparency. (d)-(f) The cross-sectional SEM views of the $MAPbBr_3$/Sa, $MAPbBr_3$/ZnO(30nm), and $MAPbBr_3$/ZnO(100nm) samples, respectively. (g) The XRD pattern of $MAPbBr_3$ nanocrystals with the corresponding Miller indexes labelled. (h) and (i) The room temperature conventional absorption (optical density - OD) and PL spectra measured for three samples identified in (d)-(f). The transitions between VB and CB of $MAPbBr_3$ and ZnO are indicated. The Stokes shift was estimated as a photon energy difference between the FE peak in absorption spectra and the PL peak.

compared, for example, to GaAs (~12 nm),[32] as will be discussed further below in details.

## III. RESULTS AND DISCUSSION

### A. Sample characterization

Figure 2(a) shows the TEM image of the as-grown colloidal cubic-shaped $MAPbBr_3$ nanocrystals together with the corresponding histogram presenting the nanocrystal size distribution which is maximized at ~19.8 ± 1.7 nm. The high-resolution TEM image [Fig. 2(b)] and XRD pattern [Fig. 2(g)] confirms the high crystallinity and 3D structure of the individual $MAPbBr_3$ nanocrystal with the typical characteristic lattice fringes spaced by ~0.42 nm.[29,33-39] Figure 2(b) also shows schematically the method of fully encapsulating $MAPbBr_3$ nanocrystals between the two sapphire plates. The fully encapsulated $MAPbBr_3$ nanocrystals demonstrate stable optical properties (at least for 4 months within which various optical measurements were performed), such as high transparency and uniform PL [Fig. 2(c)]. The thickness of $MAPbBr_3$ layer viewed by SEM is comparable to the size of nanocrystals [Fig. 2(d)-(f)], thus suggesting that no more than one layer of the closely packed $MAPbBr_3$ nanocrystals was deposited. Because 3D $MAPbX_3$ nanocrystals are known to be the basic building blocks for growing the corresponding nanoplates and nanowires, that is, 2D and 1D structures,[35,38] our samples can also be identified as quasi-2D arrays of 3D $CH_3NH_3PbBr_3$ nanocrystals to distinguish them from the 2D layered counterpart of hybrid perovskites $[(C_4H_9NH_3)_2PbBr_4]$.[38]

Figure 2(h) and (i) shows the room-temperature conventional absorption and PL spectra of the $MAPbBr_3$/Sa, $MAPbBr_3$/ZnO(30nm), and $MAPbBr_3$/ZnO(100nm) samples identified in Fig. 2(d)-(f). The absorption spectrum of the $MAPbBr_3$/Sa sample reveals two contributions associated with electronic transitions from VB to two CBs.[40,41] The ZnO layer additionally contributes to absorption spectra in the UV range for the $MAPbBr_3$/ZnO samples [Fig. 2(h)].[42] The Stokes shift was estimated as $\hbar\Delta\omega_{\text{Stokes}} = \lambda_e + \lambda_h$ = ~60 meV, where $\hbar$ is the reduced Planck constant, $\Delta\omega_{\text{Stokes}}$ is the frequency difference between the 1*s* free exciton (FE) peak in absorption spectra and the PL peak, and $\lambda_e$ and $\lambda_h$ are the corresponding reorganization energies[43] for electrons and holes, respectively. The latter quantities characterize hence the band gap renormalization appearing as the energetic difference between the unrelaxed (non-equilibrium) and relaxed (equilibrium) carriers, which can be estimated in the frame of the Fröhlich large polaron model[9,44] as $\lambda_e$ = ~32.6 meV and $\lambda_h$ = ~39.2 meV for the longitudinal-optical (LO)-phonons contribution. The intensity of the 1*s* FE absorption peak decreases in $MAPbBr_3$/ZnO due to the interfacial-electric-field-induced FE dissociation, the process which balances the relative densities of free carriers and FEs.[19] The more prominent suppression of the 1*s* FE absorption peak in the $MAPbBr_3$/ZnO(30nm) sample compared to the $MAPbBr_3$/ZnO(100nm) sample suggests that the interfacial electric field in the former is stronger than that in the latter. The exciton dissociation process is also accompanied by a blue-shift of PL-peak (~10 meV), which is greater in the $MAPbBr_3$/ZnO(30nm) sample as well [Fig. 2(i)]. These facts together with good coincidence between reorganization energies and the Stokes shift all confirm the FE nature of the band-edge light emission at room temperature. The latter statement is also well consistent with the large polaronic exciton binding energy in $MAPbBr_3$ (~35 meV),[17,23] thus exceeding substantially the room temperature $k_BT$ = 25.7 meV, where $k_B$ in the Boltzmann constant and $T$ is the temperature. We also note that because the size of $MAPbBr_3$ nanocrystals (~20 nm) substantially exceeds the exciton Bohr radius (~2.0 nm),[45] any quantum-confinement-induced effects are expected to be negligible.

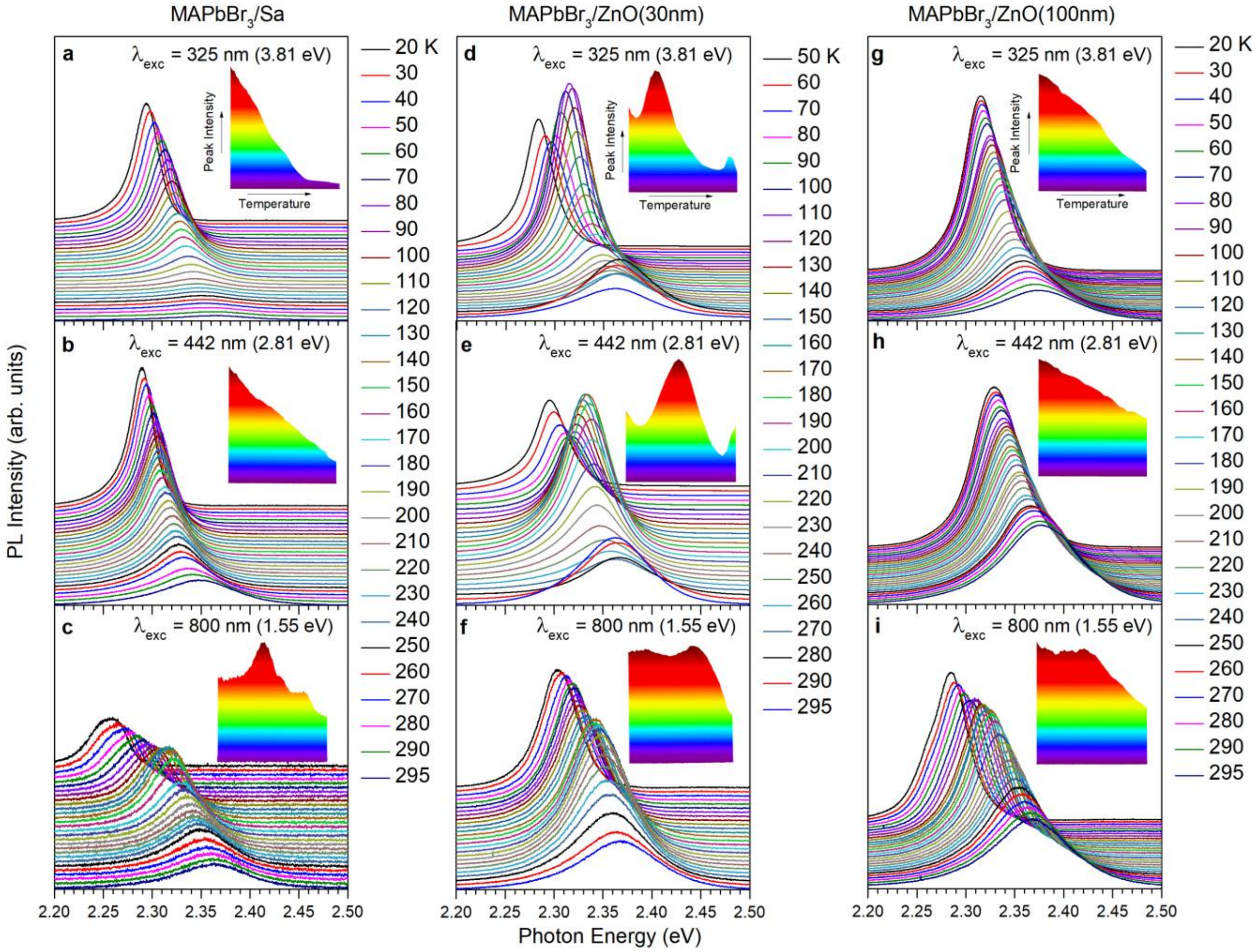

**FIG. 3** PL spectra measured for the $MAPbBr_3$/Sa (a) – (c), $MAPbBr_3$/ZnO(30nm) (d) – (f), and $MAPbBr_3$/ZnO(100nm) (g) – (i) samples at different temperatures, as indicated by the corresponding colors, and with different excitation photon energies, as indicated for each of the panels. The individual baselines for each of the spectra are arbitrary shifted for better observation. The color map graphs present the corresponding variations of the PL-peak intensity with increasing temperature.

The typical phonons contributing to the temperature-dependent dynamics include the $PbBr_6$ octahedra twist mode (TO-type) with frequency ~40 cm$^{-1}$ (~5 meV) and the distortion mode (LO/TO-type) with frequency ~58 cm$^{-1}$ (~7.2 meV).[46,47] The interaction between MA cations and $PbBr_6$ anions results in a broad MA torsional (MAT) mode peaked at ~300 cm$^{-1}$ (~37.2 meV), which governs the orientation dipole order of MA cations in the whole crystal.[46] MA internal (MAI) modes have much higher frequencies of ~900 - 3200 cm$^{-1}$ (112 - 397 meV).[46,47] However, owing to a global lattice compression in nanocrystals,[36] the frequency of the LO-phonon mode observed for $MAPbBr_3$ nanocrystals increases to ~150 cm$^{-1}$ (~18.6 meV).[48,49] Because the lattice compression varies with the nanocrystal size, TO/LO-phonon energy is expected to be spread over a few meV.[49] Consequently, we will refer further below to the following low-energy lattice vibrations: (i) the TO-phonon mode with averaged energy $\langle \hbar\omega_{TO} \rangle$ ~5.0 meV, (ii) the LO-phonon mode with averaged energy $\langle \hbar\omega_{LO} \rangle$ ~18.6 meV, (iii) the MAT-phonon modes with averaged energy $\langle \hbar\omega_{MAT} \rangle$ ranging between ~35 and ~90 meV, and (iv) the MAI-phonon modes and their combinations with averaged energy $\langle \hbar\omega_{MAI} \rangle$ ranging between ~100 and ~700 meV.

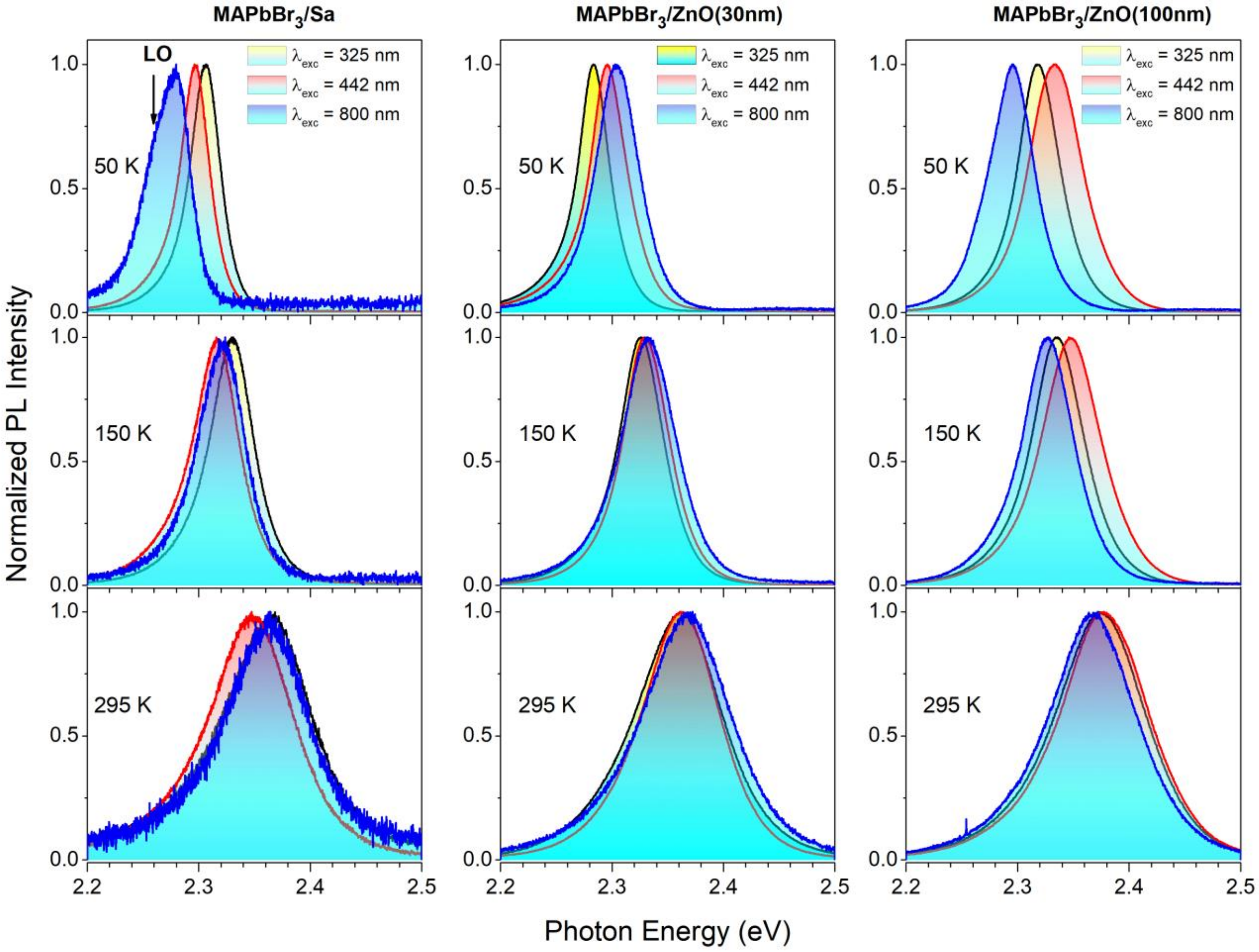

**FIG. 4** PL spectra of the three samples, as indicated on the top of each of the columns, which were measured at the different excitation wavelengths, as indicated by the corresponding colors, and at different temperatures of 50, 150, and 296 K, as indicated for each of the panels.

## B. Structural phase transitions in 3D $MAPbBr_3$ nanocrystals

The structural phase transitions in $MAPbX_3$ have been monitored for single-crystal $MAPbBr_3$[13-18,20,21] and $MAPbI_3$[16,19,20] using the XRD,[13,14,18] absorption,[16] reflection,[17] one-photon excited PL,[13,14,18,19,21] two-photon excited PL,[15] and dielectric response[20] techniques. Polycrystalline films,[21,22,23] microplate crystals[24] and nanowires[25] of $MAPbBr_3$[23] and $MAPbI_3$[21-25] have also been studied using the XRD,[22] absorption,[21,23] charge transport[24] and one-photon excited PL[21,22,24,25] techniques to recognize the structural phase transitions on the nanoscale. The structural phase transition in $MAPbX_3$ usually appears as the stepwise shift of the corresponding spectral band[13-25] whereas its intensity is less suitable for this purpose.[18,19,21,22,24] Three structural phase transitions were found to occur at $T$ ~145 K [orthorhombic(O)-to-tetragonal(T1)], at $T$ ~155 K [tetragonal(T1)-to-tetragonal(T2)], and at $T$ ~237 K [tetragonal(T2)-to-cubic(C)], which usually appear in single-crystal $MAPbX_3$ and its polycrystalline thin film.[13,15] However, even a shift of the absorption and PL bands was incapable for recognizing the structural phase transition in $MAPbX_3$ nanocrystals.[21] The reason that a stepwise shift of the absorption and PL bands is no longer observable in $MAPbX_3$ nanocrystals has been suggested to arise from the configurational entropy loss upon suppressing long-range MA polar order.[21,27,28]

To study the structural phase transition in a layer of 3D $MAPbBr_3$ nanocrystals, we measured the temperature dependences of PL from the aforementioned three samples using the three laser excitation regimes of photon energy (i) 3.81 eV ($\lambda_{exc}$ = 325 nm) being above ZnO and $MAPbBr_3$ band gaps ($E_g$= ~3.37 and ~2.3 eV, respectively); (ii) 2.81 eV ($\lambda_{exc}$ = 442 nm) being below ZnO band gap but above $MAPbBr_3$ band gap; (iii) 1.55 eV ($\lambda_{exc}$ = 800 nm) being below ZnO and $MAPbBr_3$ band gaps [Fig. 1(b)]. Figure 3 shows PL spectra measured as a function of temperature using the three different laser excitations, as indicated for each of the panels. Additionally, Figure 4 shows PL spectra measured at temperatures $T$ = 50, 150, and 285 K, which correspond to the orthorhombic, tetragonal, and cubic structural phases of single-crystal $MAPbBr_3$, respectively. All PL spectra demonstrate a characteristic ≤100 meV blue-shift with increasing temperature from $T$ ~20 to 295 K (Fig. 3 and Fig. 4). The position of PL peak in low-temperature spectra ($T$ = 50 K) slightly vary with excitation photon energy in the range of ~30 - 40 meV. This variation sets up the range of inhomogeneous broadening, which is believed to be due to the nanocrystal structural imperfectness since the $MAPbBr_3$/ZnO heterointerface does not affect significantly the position of PL bands and their full width at half maximum (FWHM). Moreover,

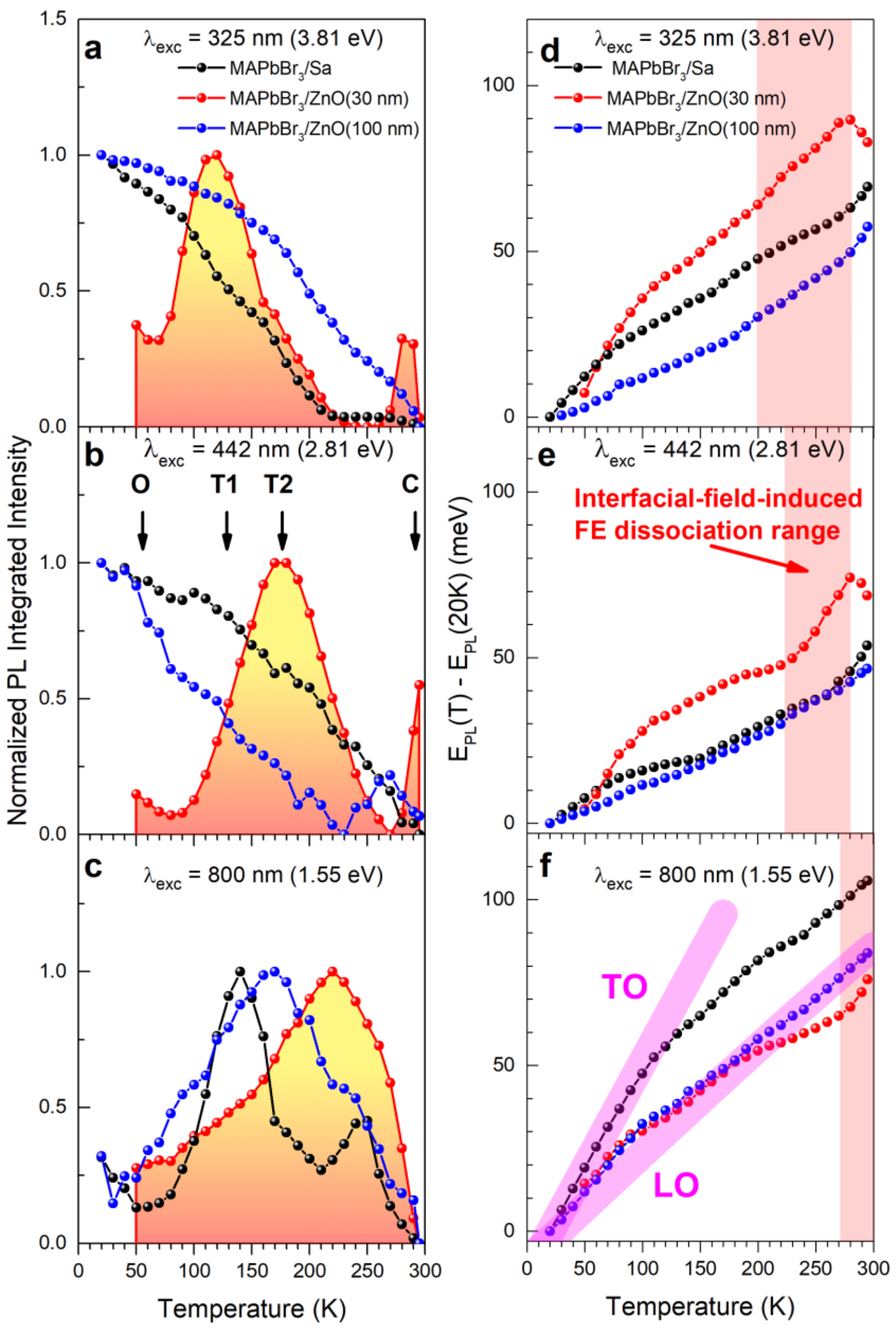

**FIG. 5** (a), (b), (c) The normalized temperature dependences of integrated PL intensities measured with the three laser excitations $\lambda_{exc}$, as indicated for each of the panels, for the three samples, as indicated by the corresponding colors. The maximal interfacial electric field sample is highlighted by a yellow-red filling. (d), (e), (f) The corresponding temperature dependences of PL peak position plotted as $E_{PL}(T) - E_{PL}(20\ K)$. The orthorhombic (O), tetragonal (T1 and T2), and cubic (C) structural phases are indicated by arrows in (b). The interfacial-field-induced FE dissociation ranges are marked in (d) – (f) as vertical light-red rectangles. The numerically simulated results obtained using Eq. (14) for TO and LO phonons are shown in (f).

the LO-phonon sideband[48] is red-shifted from PL peak by ~18 meV for the $MAPbBr_3$/Sa sample when two-photon excitation is applied ($\lambda_{exc}$ = 800 nm) (Fig. 4). As temperature increases, both the PL peak position variations and the LO-phonon sideband peak are masked by homogeneous broadening due to dominant carrier scattering with optical phonons. The bandwidth of the room-temperature PL bands (FWHM) reaches ~90 meV, indicating that homogeneous and inhomogeneous broadenings are somewhat comparable. The PL broadening dynamics with increasing temperature will be discussed in detail in the next section.

There are two general tendencies characterizing the temperature-dependent dynamics of one-photon-excited PL ($\lambda_{exc}$ = 325 nm or 442 nm). Specifically, PL peak intensities [Fig. 3(a), (b), (d), (e), (g), (h)] and PL band integrated intensities [Fig. 5(a) and (b)] both show either a monotonic decrease with increasing temperature from 20 to 295 K for the $MAPbBr_3$/Sa and $MAPbBr_3$/ZnO(100nm) samples or a non-monotonic trend maximizing at moderate temperatures for the $MAPbBr_3$/ZnO(30nm) sample. In contrast, all samples show non-monotonic trends for two-photon-excited PL ($\lambda_{exc}$ = 800 nm) [Fig. 3(c), (f), (i) and Fig. 5(c)]. As we mentioned above, all PL peak positions reveal a nearly monotonic blue-shift with increasing temperature [Fig. 5(d)-(f)]. However, this general tendency might also include an inflection point at $T$ ~80 ± 30 K, most likely pointing to the two different processes being involved. The largest blue-shift is observed for two-photon-excited PL (~100 meV), indicating that the effect depends on the photoexcited carrier density, since at least 10-fold higher carrier density was photoexcited in this case compared to one-photon-excited PL. The temperature dependences of PL peak position for the $MAPbBr_3$/ZnO(30nm) sample also demonstrate the minor features when approaching room temperatures [Fig. 5(d) - (f)], pointing to more complicated dynamics in this case. The fairly monotonic blue-shift with increasing temperature is well consistent with that reported for $MAPbI_3$ nanocrystals and suggested to provide evidence that $MAPbI_3$ nanocrystals do not undergo the bulk phase transitions.[21] Alternatively, we show that although the temperature dependences of PL peak position reveal a fairly monotonic behaviour, the temperature dependences of PL intensity can be either monotonic or non-monotonic depending on the PL excitation regime applied.

Figure 5(a)-(c) clearly demonstrates this dual behaviour for the $MAPbBr_3$/Sa sample. Specifically, although one-photon-excited PL decays with increasing temperature nearly monotonic, there are two distinct peaks for two-photon-excited PL, the positions of which ($T$ ~140 K and ~245 K) closely match those known for the orthorhombic-to-tetragonal and tetragonal-to-cubic phase transitions in single-crystal $MAPbBr_3$.[13-21] The temperature dependences of the integrated PL intensity can be fitted using the multiple Mott equation,[50] which for one-photon-excited PL, takes into consideration phonon-assisted PL quenching in all the structural phases (three terms), as well as in the phase transition regions (two terms),

$$I_{PL}(T) = \sum_{\mathrm{i}=1}^{\mathrm{i}=5} \left( \frac{I_{PLi}(0)}{1+c_i e^{-E_{ai}/k_\mathrm{B}T}} \right), \qquad (1)$$

where $I_{PLi}(0)$ is PL intensity at $T$ = 0 for each of the terms, $c_i$ is the pre-exponential factors characterizing the relative probabilities of non-radiative decay, and $E_{ai}$ is the corresponding activation energies. We note that all 5 terms in Eq. (1) are positive for one-photon-excited PL, thus characterizing the overall phonon-assisted PL quenching, while partial contributions from the specific components can be weakly recognized [Fig. 6(c)]. However, the situation changes dramatically when switching to two-photon excited PL. Consequently, phonon-assisted PL quenching in each of the structural phases (three positive terms) still contribute into the temperature-dependent dynamics, however, together with PL intensity increase when the structural phase changes towards the higher symmetry one (two negative terms) [Fig. 6(a) - (c)]. This observation demonstrates a higher sensitivity of the nonlinear absorption coefficient to the crystalline lattice symmetry and suggests that the specific phonon modes participate in the structural phase transitions[51] similarly to PL non-radiative decay.[50] Consequently, temperature dependences of both one-photon-excited and two-photon-excited PL intensities can be fitted using the same $c_i$ and $E_{ai}$ parameters, nevertheless, the intensity $I_{PLi}(0)$ of two terms governing the structural phase transition dynamics change sing when switching to two-photon-excited PL. Specifically, PL quenching in the orthorhombic phase involves TO-phonons ($E_{a1}$ ~5 meV). In contrast, PL quenching in the tetragonal/cubic phase involves MAI-phonons ($E_{a3}$ ~204 meV, $E_{a5}$ ~413 meV). The orthorhombic-to-tetragonal phase transition is a phonon-assisted process which occurs owing to MAT-phonon activation ($E_{a2}$ ~45 meV) whereas the tetragonal-to-cubic phase transition involves MAI-phonons ($E_{a4}$ ~615 meV). We note also that the orthorhombic-to-tetragonal structural phase transition in 3D $MAPbBr_3$ nanocrystals is spread out over the $T$ ~70 - 140 K range, which is believed to be due to the configurational entropy loss and the corresponding structural phase instability when, unlike in single-crystal $MAPbX_3$, free rotations of the MA ions are no longer restricted strongly by long-range polar order.[27,28] The resulting local field fluctuations in $MAPbX_3$ nanocrystals and the liquid-like motion of MA cations[1] weaken and smooth distortions of $PbX_6$ octahedra which are responsible for the band-edge electronic transitions, thus eliminating the stepwise shift of the corresponding absorption and PL bands.

Although the temperature dependences of the integrated PL intensity for the $MAPbBr_3$/ZnO(100nm) sample are quite similar to those of the $MAPbBr_3$/Sa sample, they differ significantly for the $MAPbBr_3$/ZnO(30nm) sample [Fig. 5(a)-(c)]. The dynamics can be associated with that being caused by the interfacial electric field.[19] Specifically, one should distinguish between the two principally different PL excitation regimes. One of them ($\lambda_{exc}$ = 325 nm) deals with the excitation of carriers in both $MAPbBr_3$ and ZnO. The charge separation process at the $MAPbBr_3$/ZnO heterointerface in this case is not efficient enough because although the photoexcited holes in both materials tend to reside in $MAPbBr_3$, the majority of photoexcited electrons in $MAPbBr_3$ do not leave it since the edge of the ZnO CB is filled by electrons photoexcited in ZnO [Fig. 1(b)]. The second regime involves the excitation of carriers exclusively in $MAPbBr_3$ ($\lambda_{exc}$ = 442 and 800 nm) and hence the interfacial electric field at the $MAPbBr_3$/ZnO heterointerface is formed with high efficiency since electrons can freely move to the CB of ZnO whereas holes remain in $MAPbBr_3$ [Fig. 1(b)]. One can hence vary the strength of the interfacial electric field by varying the photoexcited carrier density and $\lambda_{exc}$.

The thickness of the ZnO layer also significantly affects the interfacial electric field strength. Specifically, the interfacial electric field in the $MAPbBr_3$/ZnO(30nm) sample is expected to be much stronger compared to that in the $MAPbBr_3$/ZnO(100nm) one. This statement can be clarified in the framework of the two effects that can potentially occur at the $MAPbBr_3$/ZnO heterointerface: (i) the strain-induced effect and (ii) the charge-separation-induced effect. The first one takes into consideration that the ZnO layer was grown on the sapphire substrate and hence the thicker the ZnO layer, the stronger the residual strain should act on $MAPbBr_3$ nanocrystals.[51] The second effect also depends on the ZnO layer thickness, but in the opposite way. Because the strength of the interfacial electric field is proportional to the carrier density separated at the $MAPbBr_3$/ZnO heterointerface, the thicker the ZnO layer, the lower the carrier density in it and hence the weaker the interfacial electric field. This principal difference between the strain and electric field induced effects can be distinguished by testing two samples of different ZnO layer thicknesses, as it has been done in the current study. Specifically, the non-monotonic behaviour observed for the $MAPbBr_3$/ZnO(30nm) sample with one-photon excitation ($\lambda_{exc}$ = 325 and 442 nm) points to the stronger interfacial electric field being involved in this sample. Moreover, as the strength of the interfacial electric field increases ($\lambda_{exc}$ = 442 nm), the shift of the structural phase transition towards the higher temperature range also progresses. Alternatively, the temperature dependences of the $MAPbBr_3$/Sa and $MAPbBr_3$/ZnO(100nm) samples demonstrate a similar monotonic behaviour, suggesting that the interfacial electric field in $MAPbBr_3$/ZnO(100nm) sample is as weak as that in the $MAPbBr_3$/Sa sample and proving that the strain-induced effect is negligible [Fig. 5(a) and (b)]. This tendency is also confirmed

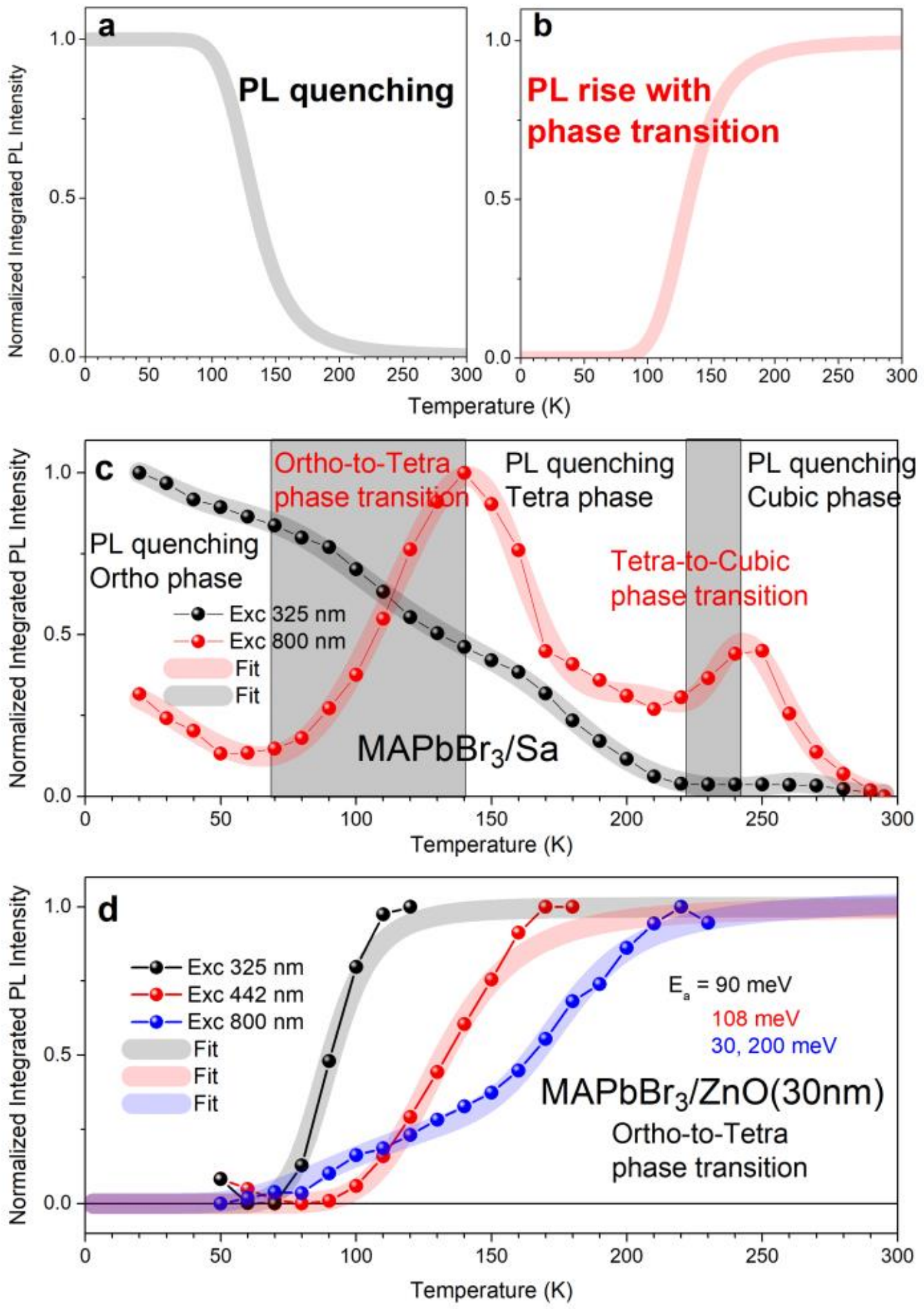

**FIG. 6** (a) An example of the phonon-assisted PL quenching. (b) An example of the phonon-assisted structural phase transition presented using the Mott equation with negative amplitude. (c) The multiple Mott equation fit of the temperature dependences of the PL integrated intensity of the $MAPbBr_3$/Sa sample measured in the one-photon and two-photon excitation regimes. The ranges of PL quenching in different structural phases and the structural phase transitions are indicated. (d) The orthorhombic-to-tetragonal structural phase transition range and its extension towards the higher temperatures for different excitation regimes in the $MAPbBr_3$/ZnO(30nm) sample, as indicated by the corresponding colors. To fit experimental curves, the Mott equation with negative amplitude was used. The corresponding activation energies are listed.

using two-photon-excited PL ($\lambda_{exc}$ = 800 nm), despite the non-monotonic temperature dependences for all the samples. Specifically, the temperature dependence of two photon-excited PL intensity for the $MAPbBr_3$/ZnO(100-nm) sample looks more like that for the $MAPbBr_3$/Sa sample [Fig. 5(c)].

The orthorhombic-to-tetragonal phase transition in the $MAPbBr_3$/ZnO(30nm) sample reaches the extremely broad temperature range of $T$ ~70 – 230 K [Fig. 5(a)-(c) and Fig. 6(d)], confirming once again that the interfacial electric field in this sample is enhanced. The dynamics also demonstrate more clearly the existence of T1 and T2 subphases. The corresponding activation energies are in the range of MAT- and MAI-phonons [Fig. 6(d)], confirming that the structural phase instability results from the MA dipole order suppression. Additionally, if the $\lambda_{exc}$ = 800 nm excitation regime is applied, the cubic structural phase feature is not observed for the $MAPbBr_3$/ZnO(30nm) sample even for temperatures ranging up to $T$ ~ 295 K [Fig. 5(a)-(c)], suggesting that the room-temperature structural phase in this case most likely is also instable, being the mixture of the orthorhombic and tetragonal phases.

To estimate how far the interfacial electric field is extended inward towards the $MAPbBr_3$ nanocrystal core, we calculated the Thomas-Fermi screening length for the photoexcited carrier densities $n_c$ = 1.0 × $10^{19}$ $cm^{-3}$, $n_c$ = 1.0 × $10^{18}$ $cm^{-3}$ and $n_c$ = 1.0 × $10^{17}$ $cm^{-3}$ (see the Experimental Methods section) as $1/k_{TF}$ = ~2.7 nm, ~3.9 nm and ~5.8 nm, respectively, with $k_{TF}$ being the Thomas-Fermi wavevector defined as $k_{TF}^2 = \left(\frac{3}{\pi^4}\right)^{1/3} \frac{m_c^* e^2 n_c^{1/3}}{\varepsilon_s \varepsilon_0 \hbar^2}$,[19,53] where $m_c^*$ is the carrier effective mass ($m_e^*$ = $0.13m_0$ and $m_h^*$ = $0.19m_0$ for electrons and holes, respectively, with $m_0$ being the free-electron mass), $e$ is the electron charge, $\varepsilon_s$ = 21.36 is the static dielectric constant, and $\varepsilon_0$ is the permittivity of free space.[9] Consequently, the Thomas-Fermi screening length naturally decreases with increasing carrier density, indicating that the interfacial electric field tends self-consistently to be confined at the heterointerface when the photoexcited carrier density increases. Because the range of the orthorhombic-to-tetragonal structural phase transition also extends with increasing photoexcited carrier density, the latter behaviour implies that the strength of the interfacial electric field mainly governs the structural phase instability in the whole $MAPbBr_3$ nanocrystal rather than the field extension inward towards the nanocrystal core. The short-range Thomas-Fermi screening length also suggests that the long-range MA dipole order which was suppressed substantially in the whole $MAPbBr_3$ nanocrystal cannot be restored by the interfacial electric field.

We also note that the temperature dependences of PL peak position observed for the $MAPbBr_3$/ZnO(30nm) sample demonstrate some additional features when approaching room temperature [Fig. 5(d) - (f)]. We associate these features with FE dissociation because of stronger interfacial electric field in this sample.[19] Specifically, the interfacial electric field dissociates FEs in the tetragonal phase, giving rise to the blue-shift of the PL band because of progressive switching from FE to band-to-band recombination. The rate of this process strongly depends on the electric field strength, thus being maximised for the $\lambda_{exc}$ = 442 nm and $\lambda_{exc}$ = 800 nm excitation regimes. Once the tetragonal-to-cubic structural phase transition occurs, the FE dissociation process weakens, giving rise to the red-shift of the PL band since FE binding energy in the cubic structural phase is higher compared to that in the tetragonal phase.[54] It should be noted that applying $\lambda_{exc}$ = 800 nm excitation, one can observe only the initial blue-shift of the PL band because the tetragonal-to-cubic structural phase transition in this case occurs at temperatures higher than room temperature [Fig. 5(f)].

### C. PL excitation mechanisms

The formation of the interfacial electric field of different strengths is also one of the key circumstances of why PL technique becomes sensitive enough to negligible structural distortions in 3D $MAPbBr_3$ nanocrystals. Specifically, this behaviour is realized because PL excitation involves the absorption rates governed by the second-order and third-order nonlinear susceptibilities, which owing to their higher rank tensor nature compared to the first-order susceptibilities, are known to demonstrate a higher spatial sensitivity to the lattice symmetry.[55-58] The situation emerging is known as electric-field-induced one-photon-excited PL and two-photon-excited PL, both involving nonlinear susceptibilities.[55] This behaviour is in stark contrast to the conventional one-photon-excited PL which loses sensitivity to the structural phase transition in 3D $MAPbX_3$ nanocrystals because of structural phase instability and the corresponding negligible distortions of $PbX_6$ octahedra responsible for light-emitting process. It is worth noting that all three structural phases in $MAPbX_3$ materials are centrosymmetric.[59,60] This statement significantly distinguishes between PL excitation regimes through the light absorption rate. Specifically, PL intensity can be expressed as[61]

$$I_{PL} \propto R_{em} \quad , \tag{2}$$

where $R_{em} \propto n_e n_h$ is the emission rate caused by carrier radiative recombination with $n_e$ and $n_h$ being the density of electrons and holes, respectively. The latter process is known as bimolecular recombination and mainly appears through electroluminescence (EL), when $n_e$ and $n_h$, in general, can be different as a consequence of the specific structure of the samples, their doping type, as well as the carrier injection level. Additionally, for highly efficient light-emitters, $n_e$ and $n_h$ should be low enough to guarantee the carrier wavevector conservation in the recombination process.[61]

In contrast, in PL experiments one always excites equal numbers of electrons and holes, $n = n_e = n_h$ (each photon with energy exceeding band gap energy excites two particles, electron and hole), with energies equal to one half of the difference between photon and band gap energies. The density of photoexcited carrier is usually much higher compared to the intrinsic carrier density (the doping level). Furthermore, PL resulting from recombination between nonthermalized (hot) carriers (hot PL) should also be negligible since the wavevector is not strictly conserved for them. However, if wavevector conservation is not necessary, the situation which may happen due to the trapping of carriers by defects or carrier interaction with phonons (including the polaron formation as well), the rate $R_{em} \propto n$ (monomolecular recombination) if the photoexcited carrier density exceeds the intrinsic carrier density.[61] The latter proportionality indicates that two-particle recombination is a highly probable process emitting a single photon. The monomolecular recombination is hence a direct opposite of the PL excitation process and its rate is known to be significantly enhanced as compared to that of bimolecular recombination.[61] Because $n \propto R_{abs} R_{rel}$, where $R_{abs}$ and $R_{rel}$ are rates of light absorption and carrier relaxation to the light-emitting states, respectively,[61] and because $R_{rel}$ is expected to be a constant for the fixed incident photon energy, as that occurs in our case, PL intensity can be expressed as

$$I_{PL} \propto R_{abs}, \tag{3}$$

that is, being predominantly governed by the absorption rate $R_{abs}$ (in units of $s^{-1}$), which, in general, is a sum of several contributions associated with one-photon absorption $R_{abs}^{(1)}$, two-photon absorption $R_{abs}^{(2)}$ if the excitation light intensity ($I_{exc}$) is strong enough, and one-photon electroabsorption $R_{eabs}^{(1)}$ if an external or internal electric field is applied, so that

$$\begin{aligned} R_{abs} &= R_{abs}^{(1)} + R_{eabs}^{(1)} + R_{abs}^{(2)} \\ &= \frac{I_{exc}}{\hbar\omega_{exc}} \left\{ \left[ \sigma_{abs}^{(1)} + \sigma_{eabs}^{(1)} \right] + \frac{I_{exc}}{\hbar\omega_{exc}} \sigma_{abs}^{(2)} \right\}, \end{aligned} \tag{4}$$

where $\hbar\omega_{exc}$ is the excitation photon energy, $\sigma_{abs}^{(1)}$, $\sigma_{eabs}^{(1)}$ and $\sigma_{abs}^{(2)}$ are the corresponding one-photon and two-photon cross sections.[55] If $\hbar\omega_{exc} > E_g$ , then the $R_{abs}^{(1)}$ and $R_{eabs}^{(1)}$ terms dominate the PL excitation dynamics. On the contrary, if $\hbar\omega_{exc} < E_g$ , then the $R_{abs}^{(2)}$ term completely governs the PL excitation mechanism. Consequently, the following proportionalities $I_{PL} \sim I_{exc}$ and $I_{PL} \sim I_{exc}^2$ correspond to one-photon-excited and two-photon-excited PL, respectively.[62-65] These relations can be confirmed experimentally when analyzing the slope of the power dependences of $I_{PL}$ presented in a log-log plot (Fig. 7).

However, the absorption rates of the one-photon and two-photon absorption processes are known to be proportional to the imaginary part of the first-order and third-order optical susceptibilities ( $\sigma_{abs}^{(1)} \sim Im[\chi^{(1)}]$ and $\sigma_{abs}^{(2)} \sim Im[\chi^{(3)}]$ ), respectively.[55] Consequently, this approach is well consistent with the aforementioned centrosymmetry restriction applied to $MAPbBr_3$ crystals, according to which the second-order nonlinear process [ $\chi^{(2)}$ ] is not allowed, whereas the linear [$\chi^{(1)}(-\omega;\omega)$] and third-order nonlinear [$\chi^{(3)}(-\omega;\omega,\omega,-\omega)$] processes should completely govern the one-photon and two-

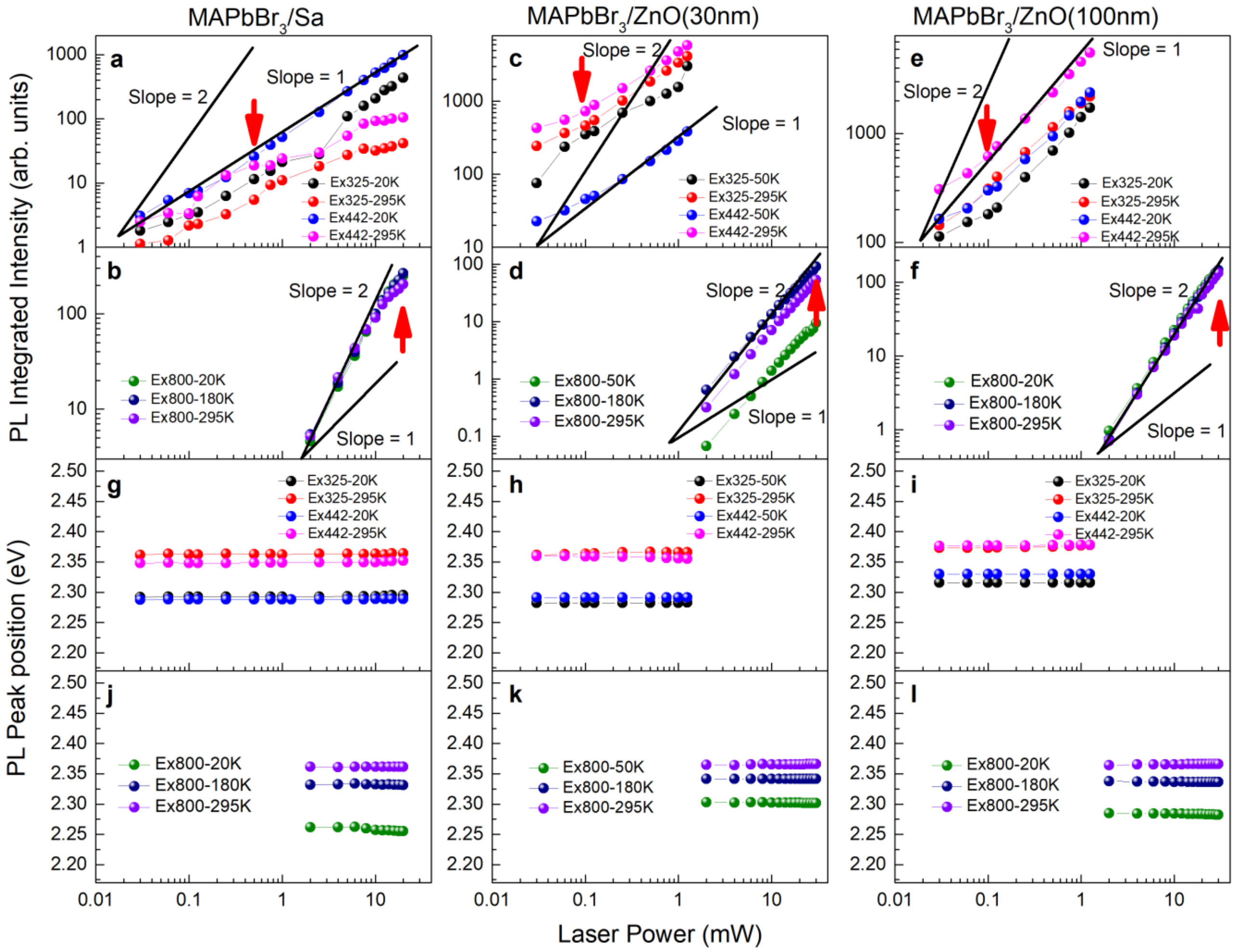

**FIG. 7** (a)-(f) Power dependences of the integrated PL intensity presented in the log-log scale for the three samples, as indicated at the top, which were measured at different temperatures and excitations, as indicated by the corresponding colors. The slopes of the rise of the integrated PL intensity with increasing laser power indicate the one-photon and two-photon excitation regimes. The red vertical arrows mark average powers at which the temperature-dependent PL measurements were performed. (g)-(l) Power dependences of PL-peak position presented in the semi-log scale for the same three samples, which demonstrate no significant effect on the PL-peak position, confirming hence high resistance of a fully encapsulated layer of 3D $MAPbBr_3$ nanocrystals against light exposure.

photon absorption in these materials, appearing through one-photon-excited and two-photon-excited PL, respectively. However, once the crystalline lattice is getting distorted by an external or internal electric field, for example, the centrosymmetry breaking allows the second-order nonlinear process to appear through the linear electro-optic effect $\sigma^{(1)}_{eabs} \sim Im[\chi^{(2)}(-\omega;\omega,0)]$ (Franz-Keldysh effect).[55-58,61] Additionally, the electric field can induce a quadratic electro-optic effect $\sigma^{(1)}_{eabs} \sim Im[\chi^{(3)}(-\omega;\omega,0,0)]$ (Kerr effect).[55-58] We note that PL excitation involving $\chi^{(2)}(-\omega;\omega,0)$ and $\chi^{(3)}(-\omega;\omega,0,0)$ still produces a one-photon-excited PL response, although the light absorption rates are characterized by nonlinear electro-optic susceptibilities. This brief discussion of nonlinear optics highlights an advantage of the electric-field-induced one-photon-excited and two-photon-excited PL for monitoring structural phase transitions in hybrid perovskite nanoscale materials. This behaviour results from the fact that these techniques exploit the higher sensitivity of nonlinear optical and electro-optical susceptibilities to the crystalline lattice distortions compared to the conventional linear optical processes.[55-58] Specifically, $\chi^{(1)}(-\omega;\omega)$ is a second rank tensor containing 9 elements, whereas $\chi^{(2)}(-\omega;\omega,0)$ and $\chi^{(3)}(-\omega;\omega,0,0)$, $\chi^{(3)}(-\omega;\omega,\omega,-\omega)$ are the third and fourth rank tensors containing 27 and 81 elements, respectively.[55] This principal difference also indicates that both $R^{(1)}_{eabs}$ and $R^{(2)}_{abs}$ mainly characterize PL excitation in the nanocrystal core, contrary to $R^{(1)}_{abs}$ which characterizes PL excitation in both the nanocrystal core and the nanocrystal surface. Consequently, because the ratio of the surface-to-core PL contributions for nanocrystals is large enough and because the surface states are less sensitive to the structural phase transitions in the core, the structural phase transitions in $MAPbX_3$ nanocrystals can be significantly masked by PL from the surface states when conventional one-photon excitation is applied. This circumstance together with the structural phase instability occurring within the broad temperature range seems to be a reason why the structural phase transitions in $MAPbX_3$ nanocrystals have never been observed.

It is worth noting that PL peak position remains almost unchanged with increasing laser power for all the samples and

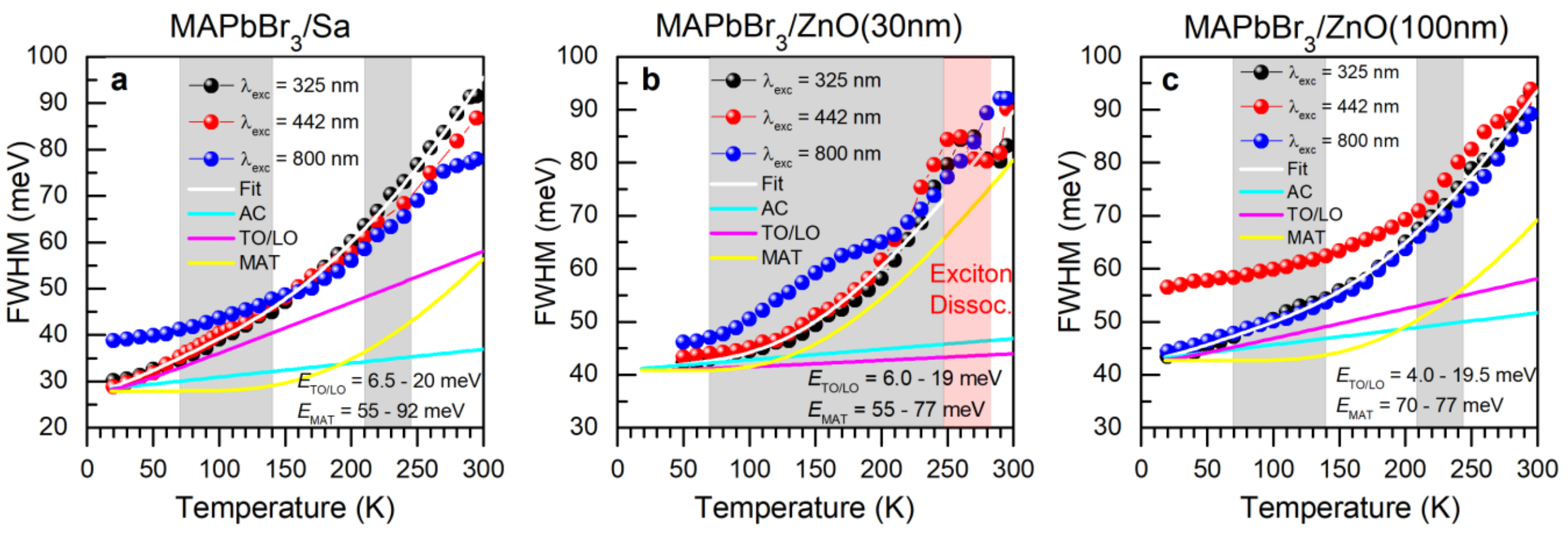

**FIG. 8** (a)-(c) The temperature dependences of the PL band FWHM measured at the three laser excitations $\lambda_{exc}$, as indicated by the corresponding colors. An example of the fit using Eq. (5) is shown for PL measured with $\lambda_{exc}$ = 325 nm, where acoustic phonon (AC), TO/LO-phonon, and MAT-phonon modes were taken into consideration. The corresponding components of the fit and the estimated ranges of the averaged phonon energies are shown. The light-gray color filled rectangles show the structural phase transition ranges defined in Fig. 6(c) and (d). The light-red color filled rectangle in (b) show the FE dissociation range for $MAPbBr_3$/ZnO(30nm) when $\lambda_{exc}$ = 325 nm or $\lambda_{exc}$ = 442 nm excitation is applied (narrowing PL band).

all the structural phases [Fig. 7(g)-(l)], thus eliminating from consideration any laser-induced sample modification effects.

### D. PL broadening dynamics

To gain deeper understanding through the phonon-assisted structural phase transitions in 3D $MAPbBr_3$ nanocrystals, the temperature dependences of the PL band FWHM for all the samples were analysed (Fig. 8). All the dependences clearly demonstrate the two-stage homogeneous broadening process appearing at low ($T$ ~20 - 140 K) and moderate ($T$ ~140 - 295 K) temperatures. These two temperature intervals closely match those corresponding to the orthorhombic and tetragonal-cubic phases in single-crystal $MAPbBr_3$, respectively. However, PL broadening dynamics with temperature is expected to be governed rather by various phonons being involved than the structural phase transitions. The temperature dependences of the PL band FWHM also reveal additional features for the $MAPbBr_3$/ZnO(30nm) sample when approaching room temperature. These features are due to the interfacial-field-induced FE dissociation, which additionally to the blue-shift of the PL band [Fig. 5(d) – (f)] is accompanied by its narrowing.[19] Further broadening of the PL band with increasing temperature occurs when the tetragonal phase transforms to the cubic one, at which FE binding energy is increased and hence the FE dissociation process is slowed down, as discussed in the preceding section for the PL peak position dynamics.

To analyze the PL band FWHM variations with temperature, we use a phenomenological approximation for phonon-induced broadening[19,66,67]

$$\gamma = \gamma_0 + \gamma_{ac}T + \frac{\gamma_{\mathrm{TO/LO}}}{\exp\left(\frac{\langle\hbar\omega_{\mathrm{TO/LO}}\rangle}{k_{\mathrm{B}}T}\right)-1} + \frac{\gamma_{\mathrm{MAT}}}{\exp\left(\frac{\langle\hbar\omega_{\mathrm{MAT}}\rangle}{k_{\mathrm{B}}T}\right)-1}, \qquad (5)$$

where $\gamma_0$ is inhomogeneous broadening, $\gamma_{ac}$ is the electron (hole)-acoustic-phonon coupling strength, $\gamma_{TO/LO}$ is the electron (hole)-TO/LO-phonon coupling strength, $\gamma_{MAT}$ is the electron (hole)-MAT-phonon coupling strength. Consequently, the electron (hole)-acoustic-phonon coupling strength ($\gamma_{ac}$ ~$3\times10^{-5}$ eVK$^{-1}$) is negligible [Fig. 8(a)-(c)], being of the same order as that previously reported.[19,66] The first stage of PL band broadening is due to the scattering of carriers with TO/LO-phonons, whereas the second stage can be attributed to the MAT-phonons effect. It should be especially stressed here that the MAT-phonon contribution becomes significantly enhanced for the $MAPbBr_3$/ZnO(30nm) sample [Fig. 8(b)]. This behaviour confirms the stronger interfacial electric field in this sample to occur and its significant effect on the suppression of the MA cation dipole order in the whole nanocrystal.[45] We note that the fits are not necessarily unique, thus allowing one to determine the effective energy ranges of TO/LO-phonons as $\langle\hbar\omega_{\mathrm{LO/TO}}\rangle$ ~4.0 - 17 meV and MAT-phonons as $\langle\hbar\omega_{\mathrm{MAT}}\rangle$ ~55 - 92 meV, which well match those discussed above in the sample characterization section.

We also found that the electron (hole)-MAT-phonon coupling strength ($\gamma_{MAT}$ ~245 - 679 meV) is ~100-fold greater than the electron (hole)-TO/LO-phonon coupling strengths ($\gamma_{TO/LO}$ ~2.5 - 8.4 meV). This strong coupling of electrons (holes) to MAT-phonons highlights the main specific feature distinguishing the carrier relaxation in $MAPbX_3$ compared to conventional semiconductors. Specifically, the photoexcited carriers relax down not only through the TO/LO-phonon cascade, but also through the MAT-phonon excitation. This behaviour is the reason why TO/LO-phonon bottleneck occurs in $MAPbX_3$. Specifically, because of the ultralow thermal conductivity between the sublattices, the organic sublattice heated during carrier relaxation keeps TO/LO-phonons in the inorganic sublattice at the temperature of the former, thus blocking their decay through acoustic phonons (Klemens-Ridley anharmonic process) and allowing carriers to reabsorb TO/LO-phonons.[2-4] This process is expected to be enhanced in nanocrystals as a consequence the additional reduction of thermal conductivity through the nanocrystal boundaries. Consequently, the nearly monotonic blue shift of PL band with increasing temperature seems to result rather from the heating effect under TO/LO-phonon bottleneck than that being induced by a progressive distortion of $PbX_6$ octahedra. This conclusion is also well consistent with the structural phase instability in

MAPbX$_3$ nanocrystals due to the configurational entropy loss.[21,27,28]

**E. PL mechanism**

PL mechanism in MAPbX$_3$ is not trivial mainly due to a polar crystal lattice and the ultralow thermal conductivity between the sublattices. The observed dominant PL blue-shift with increasing lattice temperature ($T$) [Fig. 5(d)-(f)] is opposite to that usually predicting the band gap ( $E_g$ ) variation in conventional semiconductors[61,68,69]

$$E_g(T) = a - b\left[1 + \frac{2}{\exp\left(\frac{\theta}{T}\right) - 1}\right], \quad (6)$$

where $a$ and $b$ are fitting parameters and $\theta$ is the mean temperature of phonons taking part in the scattering process with carriers. It has recently been suggested that $E_g(T)$ can show either a decrease (red-shift) or an increase (blue-shift) depending on whether derivative $d\theta/dT$ (slope) is positive (phonon emission) or negative (phonon reabsorption), respectively.[18] The latter behavior points to the non-equilibrium dynamics, being equivalent to the introduction of the negative absolute temperature.[70]

To adapt this situation to TO/LO-phonon bottleneck, we consider the Bose–Einstein phonon occupation numbers for spontaneous TO/LO-phonon emission

$$n_{em} = \left[exp\left(\langle\hbar\omega_{\text{TO/LO}}\rangle / k_{\text{B}}T\right) - 1\right]^{-1} \quad (7)$$

and for TO/LO-phonon reabsorption

$$n_{ra} = \left\{exp\left[\langle\hbar\omega_{\text{TO/LO}}\rangle / k_{\text{B}}(T_b - T)\right] - 1\right\}^{-1}, \quad (8)$$

where $T_b$ denote the temperature at which TO/LO-phonon bottleneck occurs. This approach implies that upon photoexcitation, electrons and holes (also FEs) cool down through the TO/LO/MAT-phonon cascade. Consequently, free carriers and FEs relax down at least within a few ps timescale and their temperature in the light-emitting states prior to emission ($T_e$) is determined by the decay of TO/LO-phonons in the inorganic sublattice through the Klemens-Ridley anharmonic process[2-4,30,31], which however is controlled by the organic sublattice temperature ($T_b$). Because the further cooling of the organic sublattice through acoustic phonons is slower than that of the inorganic sublattice due to the more energetic optical phonons involved in the organic sublattice, TO/LO-phonon bottleneck in the inorganic sublattice occurs at $T_e = T_b > T$ and allows carriers in the latter to reabsorb TO/LO-phonons. The effect progresses with increasing $T$, since the organic sublattice cooling rate is reduced. Consequently, TO-phonon bottleneck occurs predominantly in the orthorhombic phase whereas LO-phonon bottleneck dominates in the tetragonal/cubic phase. The resulting TO/LO-phonon-dressing process hence lowers the electron (hole, exciton) energies by the polaron (reorganization) energy[1,7,10,71-73]. Consequently, the stronger the electron (hole, exciton)-phonon coupling, the larger the number of TO/LO-phonons contribute to the polaronic effect.

According to this model, the resulting polaronic electron (*pe*), polaronic hole (*ph*) and polaronic exciton (*PE*) quasiparticles are involved into their further recombination in MAPbBr$_3$ nanocrystals, thus completely governing their PL and transport properties. Moreover, upon TO/LO-phonon bottleneck, polaronic quasiparticles can reabsorb TO/LO-phonons to form the TO/LO-phonon vibrationally excited polaronic quasiparticles with reduced ground-state polaron energy. Consequently, polaronic quasiparticle recombination may occur in either the ground or vibrationally excited polaron states [Fig. 9(a)]. This behavior gives rise to the ~100 meV blue-shift of PL-peak with increasing temperature, unlike a red-shift in conventional semiconductors. Because the blue-shift is observed for the entire temperature range applied, one can assume that $T_b$ > 300 K whereas it should apparently be less than the material melting point of $T_b$ < 450 K. Owing to screening from other carriers and defects, polaronic quasiparticles possess a very small recombination rate (PL decay-time is very long), thus being expected to demonstrate high mobility and long-range diffusion.

To treat the experimental results, we first consider the single electron (hole) polaron energy. The polaronic band gap renormalization is known to narrow the band gap by reorganization energy, introducing the ground-state Fröhlich polaron energy for electrons and holes which can be given as[71,74]

$$\lambda_{e,h} = \langle\hbar\omega_{\text{TO/LO}}\rangle\langle\alpha_{e,h}\rangle, \quad (9)$$

where $\langle\alpha_{e,h}\rangle = \frac{e^2}{\hbar}\frac{1}{4\pi\varepsilon_0}\sqrt{\frac{m^*_{e,h}}{2\langle\hbar\omega_{\text{TO/LO}}\rangle}}\left(\frac{1}{\varepsilon_\infty} - \frac{1}{\varepsilon_s}\right)$ is the *pe*/*ph* coupling coefficient which is a measure of electron(hole)-phonon coupling strength[9,74,75]. Using $\langle\hbar\omega_{TO}\rangle$ = 5 meV and $\langle\hbar\omega_{LO}\rangle$ = 18.6 meV, one can obtain the following Fröhlich polaron coupling coefficients $\langle\alpha_e\rangle$ = 3.37, $\langle\alpha_h\rangle$ = 4.07 and $\langle\alpha_e\rangle$ = 1.75, $\langle\alpha_h\rangle$ = 2.11 for TO and LO phonons, respectively, which are well consistent with those calculated using the Feynman-Osaka model[9]. The ground-state Fröhlich polaron energy for electrons and holes is hence temperature independent and can be calculated as $\lambda_e$ = ~16.9 meV and $\lambda_h$ = ~20.35 meV for TO-phonons and $\lambda_e$ = ~32.6 meV and $\lambda_h$ = ~39.2 meV for LO-phonons. These estimates imply that LO-phonons might govern the room-temperature ~60 meV Stokes shift ($\lambda_e + \lambda_h$) discussed above in the sample characterization section. Specifically, assuming that the absorption and PL spectra manifest the unperturbed and BGR-induced dynamics, respectively, the corresponding band gap narrowing is

$$E_g^{BGR} = E_g - (\lambda_e + \lambda_h) \quad (10)$$

and the Stokes shift is hence equal to $\lambda_e + \lambda_h$. To introduce the temperature effect into the dynamics, we use the Bose–Einstein phonon occupation numbers defined above, so that

$$\begin{aligned}\lambda_{e,h} &= \langle\hbar\omega_{\text{TO/LO}}\rangle\langle\alpha_{e,h}\rangle[n_{em} + n_{ra}] \\ &= \frac{1}{2}\langle\hbar\omega_{\text{TO/LO}}\rangle\langle\alpha_{e,h}\rangle\left\{\coth\left(\frac{\langle\hbar\omega_{\text{TO/LO}}\rangle}{2k_{\text{B}}T}\right) + \coth\left[\frac{\langle\hbar\omega_{\text{TO/LO}}\rangle}{2k_{\text{B}}(T_b - T)}\right] - 2\right\}. \quad (11)\end{aligned}$$

The corresponding variation of the band gap afterwards is

$$\begin{aligned}E_g^{BGR}(T) &= E_g(0) - \frac{1}{2}\langle\hbar\omega_{\text{TO/LO}}\rangle(\langle\alpha_e\rangle + \langle\alpha_h\rangle) \\ &\times\left\{\coth\left(\frac{\langle\hbar\omega_{\text{TO/LO}}\rangle}{2k_{\text{B}}T}\right) + \coth\left[\frac{\langle\hbar\omega_{\text{TO/LO}}\rangle}{2k_{\text{B}}(T_b - T)}\right] - 2\right\} \quad (12)\end{aligned}$$

We note that when neglecting TO/LO-phonon reabsorption term ( $n_{ra} = 0$ ), Eq. (12) is reduced to the standard hyperbolic cotangent equation derived on the basis of simple thermodynamics to replace the semi-empirical Varshni relation,[76]

$$E_g^{BGR}(T) = E_g(0) - \frac{1}{2}\langle\hbar\omega_{\mathrm{TO/LO}}\rangle(\langle\alpha_e\rangle + \langle\alpha_h\rangle) \times \left\{\coth\left(\frac{\langle\hbar\omega_{\mathrm{TO/LO}}\rangle}{2k_{\mathrm{B}}T}\right) - 1\right\}. \quad (13)$$

Alternatively, when neglecting spontaneous TO/LO-phonon emission term ($n_{em} = 0$), Eq. (12) describes an increase of the band gap with temperature

$$E_g^{BGR}(T) = E_g(0) - \frac{1}{2}\langle\hbar\omega_{\mathrm{TO/LO}}\rangle(\langle\alpha_e\rangle + \langle\alpha_h\rangle) \times \left\{\coth\left[\frac{\langle\hbar\omega_{\mathrm{TO/LO}}\rangle}{2k_{\mathrm{B}}(T_b - T)}\right] - 1\right\}. \quad (14)$$

We note that Eq. (12) completely describes the variation of the polaronic band gap energy of $MAPbX_3$ nanocrystals, depending on whether TO/LO-phonon bottleneck occurs. To use this approach to *PE*s, one should consider the *PE* binding energy $\varepsilon_{PE} \equiv E_{g.s.} - \lambda_e - \lambda_h$, where $E_{g.s.}$ is the *PE* ground-state energy [Fig. 9(a)].[77] Because $\varepsilon_{PE}$ for $MAPbBr_3$ is similar to the exciton binding energy ($\varepsilon_{PE} \approx \varepsilon_{ex}$) and because $\varepsilon_{PE}$ = 35 meV is higher than the room temperature $k_BT$ = 25.7 meV,[16,77] we consider *PE*s as those dominantly contributing to PL in the temperature range of 20 - 295 K. Consequently, the exciton peak energy in absorption and PL spectra varies as follows

$$E_{ex}(T_L) = E_g(0) - E_{g.s.} = E_g(0) - \varepsilon_{PE} - \frac{1}{2}\langle\hbar\omega_{\mathrm{TO/LO}}\rangle(\langle\alpha_e\rangle + \langle\alpha_h\rangle) \times \left\{\coth\left(\frac{\langle\hbar\omega_{\mathrm{TO/LO}}\rangle}{2k_{\mathrm{B}}T}\right) + \coth\left[\frac{\langle\hbar\omega_{\mathrm{TO/LO}}\rangle}{2k_{\mathrm{B}}(T_b - T)}\right] - 2\right\}. \quad (15)$$

Accordingly, the $\coth\left[\frac{\langle\hbar\omega_{\mathrm{TO/LO}}\rangle}{2k_{\mathrm{B}}(T_b - T)}\right]$ term characterizing the non-equilibrium dynamics and governing the blue shift with temperature till $T \ll T_b$ dominates over the $\coth\left(\frac{\langle\hbar\omega_{\mathrm{TO/LO}}\rangle}{2k_{\mathrm{B}}T}\right)$ term dealing with the equilibrium relaxation dynamics and governing the PL peak red shift. As a result, the *PE* absorption and PL peaks both tend to blue shift with temperature in the broad temperature range of $T$ = 20 – 100 K, indicating a strong TO/LO bottleneck effect to occur. However, this general trend is gradually weakened for the PL peak when temperature approaches to that at which the crystalline lattice is getting unfrozen enough to initiate the anharmonic three-phonon TO/LO-phonon decay process involving acoustic phonon branches (Klemens/Ridley process). The resulting deviation of the PL peak temperature dependence from that of the *PE* absorption peak is hence coming from the weakening of TO/LO-phonon bottleneck, which mainly appears for the PL peak since *PE* relaxation (cooling) towards their ground state is required prior to light emission. Alternatively, the *PE* absorption peak temperature dependence mainly reflects the non-equilibrium dynamics, which ignores any relaxation processes. The resulting Stokes shift progresses with increasing temperature, reaching ~60 meV at room temperature. the value which presents the energetic difference between the polaronic band gap and *PE* ground-state energy. To verify whether Eq. (12) is relevant to the experimental observations, we re-plotted the temperature dependences of PL-peak position as $E_{PL}^{PE}(T) - E_{PL}^{PE}(20\ K)$, where the lowest temperature data taken at $T_L$ = 20 K is applied instead of that at $T_L$ = 0 K (Fig. 5(d)-(f)),

Alternatively, neglecting the spontaneous TO/LO-phonon emission, Eq. (12) describes the PL-peak energy increase with increasing temperature [Fig. 9(a)]. Figure 5(f) confirms the latter tendency by numerically simulated results under $T_b$ = 450 K obtained without any fitting parameters. One can clearly see that TO-phonon bottleneck dominates in the orthorhombic phase whereas LO-phonon bottleneck controls the dynamics in the tetragonal/cubic phase. The inflection point is hence a signature of switching between these two regimes.

It is worth noting that the band gap modification energy $E_g^{BGR}(T) - E_g(0)$ varies in the range of ~60 - 100 meV, which closely matches the polaronic band-edge energy $(\lambda_e + \lambda_h)$ ~ 70 meV for LO-phonons, thus suggesting that the blue shift of PL band can be associated with the LO-phonon vibrationally excited polaronic quasiparticles. We note that this mechanism, which is applied to a layer of $MAPbBr_3$ nanocrystals clearly demonstrating a single PL peak, is completely different to that proposed for thick films demonstrating dual emission features like in bulk single-crystals.[19,78] Consequently, the proposed mechanism was associated with the thermal expansion of the lattice,[78] which seems to be irrelevant for nanocrystals where the lattice is flexible enough due to structure phase instability. Furthermore, because $\lambda_e$ and $\lambda_h$ for LO-phonons are of the same order as Rashba energies ($E_R$ ~40 meV),[79] the polaronic nature of the edge states in $MAPbX_3$ materials at room temperature should dominate over that associated with the Rashba effect.

To recognize TO/LO-phonon bottleneck on the *pe*/*ph* and *PE* masses, we consider again the process of TO/LO-phonon emission/reabsorption by hot carriers. LO-phonon emission influences the *pe*/*ph* masses as $m_{pe,ph}^* = m_{e,h}^*\left[1 + \frac{\langle\alpha_{e,h}\rangle(n+1)}{6}\right]$ [74]. Accordingly, the whole dynamics can be expressed as

$$m_{pe,ph,PE}^* = m_{e,h,\mu}^*\left[1 + \frac{\langle\alpha_{e,h}\rangle(n+1)}{6} + \frac{\langle\alpha_{e,h}\rangle n_{ne}}{6}\right] = m_{e,h,\mu}^* \times \left\{1 + \frac{\langle\alpha_{e,h,\mu}\rangle}{12}\left[\coth\left(\frac{\langle\hbar\omega_{\mathrm{TO/LO}}\rangle}{2k_{\mathrm{B}}T_L}\right) + \coth\left[\frac{\langle\hbar\omega_{\mathrm{TO/LO}}\rangle}{2k_{\mathrm{B}}(T_b - T_L)}\right]\right]\right\}, \quad (16)$$

where $m_\mu^* = \frac{m_e^* m_h^*}{m_e^* + m_h^*} = 0.077m_0$ is the reduced exciton mass and $\langle\alpha_\mu\rangle = \frac{e^2}{\hbar}\frac{1}{4\pi\varepsilon_0}\sqrt{\frac{m_\mu^*}{2\langle\hbar\omega_{\mathrm{TO/LO}}\rangle}}\left(\frac{1}{\varepsilon_\infty} - \frac{1}{\varepsilon_S}\right)$ = 2.61 (1.35) is the *PE* coupling coefficient for TO (LO)-phonons. Figures 9(b) and (c) show the numerically simulated results which point out that if TO/LO-phonon bottleneck is neglected ($n_{ne}$ = 0), the polaron masses increase with increasing temperature in all the structural phases, that is, the electron (hole, exciton)-phonon coupling is enhanced. TO-phonon bottleneck significantly increases the polaron masses at $T_L$ = 0 K followed by their slight decrease with increasing temperature. This behavior indicates that the polaronic quasiparticles are strongly localized in the orthorhombic phase. Alternatively, LO-phonon bottleneck dominating in the tetragonal/cubic phase decreases the polaron masses, making them be almost temperature independent. We note that the *pe*/*ph* masses in the latter case are only slightly above the electron and hole effective masses, whereas the *PE* mass is less than those.

The effect of TO/LO-phonon bottleneck on the polaron radii can be considered using the polaron radii of Fröhlich polarons at $T_L$ = 0 K[71],

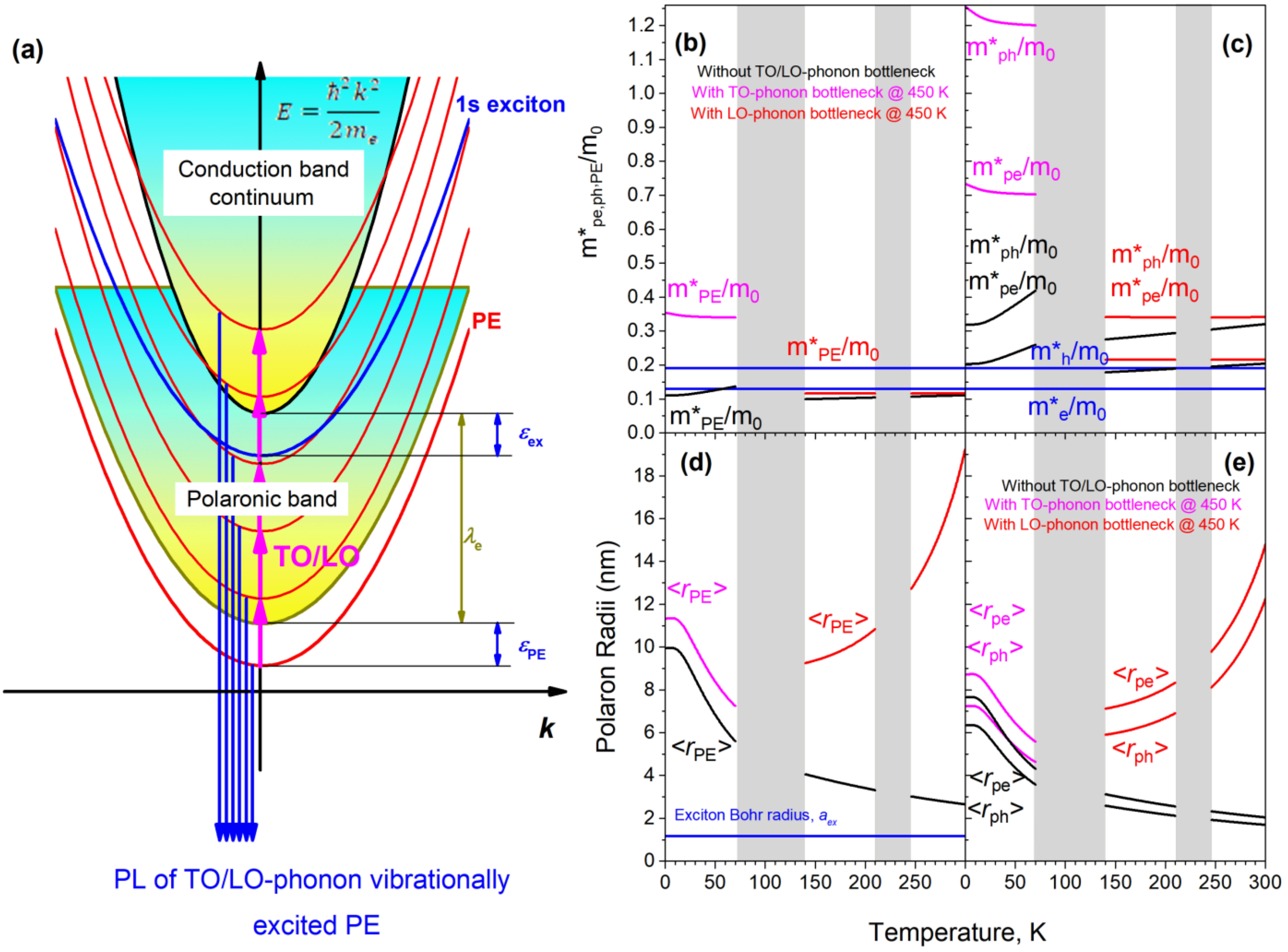

**Fig. 9 | PL mechanism and polaron masses and radii.** (**a**) A schematic presentation of CB continuum, 1*s* exciton, polaronic band, and *PE* in a layer of 3D $MAPbBr_3$ nanocrystals. A set of red curves shows a decrease of the initial *PE* energy ($E_{g.s.}$) upon TO/LO-phonon bottleneck with increasing temperature. (**b**)-(**e**) Numerical modeling of the polaron masses [(**b**) and (**c**)] and the polaron radii [(**d**) and (**e**)] for all the structural phases defined in Fig. 4 without and with TO/LO-phonon bottleneck under $T_b$ = 450 K, as indicated by the corresponding colors. The electron and hole effective masses, as well as the exciton Bohr radius are shown in blue.

$$\langle r_{pe,ph,PE}(0)\rangle = \frac{\hbar}{\sqrt{2m^*_{e,h,\mu}\langle\hbar\omega_{\text{TO/LO}}\rangle}}. \quad (17)$$

We use the aforementioned parameters to obtain the following equilibrium polaron radii at $T_L$ = 0 K when TO-phonons are involved $\langle r_{pe}\rangle$ = 7.66 nm, $\langle r_{ph}\rangle$ = 6.35 nm, $\langle r_{PE}\rangle$ = 9.95 nm and when LO-phonons are involved $\langle r_{pe}\rangle$ = 3.97 nm, $\langle r_{ph}\rangle$ = 3.29 nm, and $\langle r_{PE}\rangle$ = 5.16 nm. These values well match those calculated using the Feynman-Osaka model[9]. The polaron radii are at least three times greater than the exciton Bohr radius[80] $a_{ex} = \frac{\hbar^2\varepsilon_0\varepsilon_s}{\mu e^2}$ = 1.17 nm, which in turn is about twice the lattice constant (~0.59 nm)[9]. These estimates prove the large polaron nature (Fröhlich polaron)[71] of quasiparticles in $MAPbBr_3$ and imply that the lattice distortions spread over many lattice sites, thus allowing Fröhlich polarons to travel through the lattice as free quasiparticles. We note that the *PE* radius $\langle r_{PE}(0)\rangle$ only slightly greater than $\langle r_{pe,ph}(0)\rangle$ because the overlapped polarization clouds of *pe* and *ph* partially cancel each other.[81]

The temperature effect on the polaron radii can be treated following the general consideration for polaronic quasiparticles in quantum dots[81] and using the temperature-introducing approach[74]as

$$\langle r_{pe,ph,PE}(T_L)\rangle = \frac{1}{4\pi\varepsilon_0}\frac{e^2}{2\langle\alpha_{e,h,\mu}\rangle(n+1)\langle\hbar\omega_{\text{TO/LO}}\rangle}\left(\frac{1}{\varepsilon_\infty}-\frac{1}{\varepsilon_s}\right) = \frac{2\langle r_{pe,ph,PE}(0)\rangle}{\coth\left(\frac{\langle\hbar\omega_{\text{TO/LO}}\rangle}{2k_BT_L}\right)+1}. \quad (18)$$

Subsequently, the TO/LO-phonon bottleneck effect on the polaron radii can be expressed as

$$\langle r_{pe,ph,PE}(T_L)\rangle = \frac{1}{4\pi\varepsilon_0}\frac{e^2}{2\langle\alpha_{e,h,\mu}\rangle\langle\hbar\omega_{\text{TO/LO}}\rangle}$$

$$\times\left(\frac{1}{\varepsilon_\infty}-\frac{1}{\varepsilon_s}\right)\left(\frac{1}{n+1}+\frac{1}{n_{ne}}\right)$$

$$= 2\langle r_{pe,ph,PE}(0)\rangle\left\{\frac{1}{\coth\left(\frac{\langle\hbar\omega_{\mathrm{TO/LO}}\rangle}{2k_{\mathrm{B}}T_L}\right)+1}+\frac{1}{\coth\left[\frac{\langle\hbar\omega_{\mathrm{TO/LO}}\rangle}{2k_{\mathrm{B}}(T_b-T_L)}\right]-1}\right\}. \quad (19)$$

Figures 9(d) and (e) show the corresponding numerical modeling. One can clearly see that without TO/LO-phonon bottleneck [Eq. (18)], the polaron radii shorten with increasing temperature in all the structural phases, thus agreeing with the corresponding increase of the polaron masses.[71] TO-phonon bottleneck slightly increases the polaron radii at $T_L$ = 0 K [Eq. (19)], nevertheless, they significantly decrease with increasing temperature in the orthorhombic phase. Once LO-phonon bottleneck begins contributing to the dynamics in the tetragonal/cubic phase, the polaron radii become longer and significantly elongate with increasing temperature. The latter behavior demonstrates the weakening of the electron (hole, exciton)-phonon coupling. The resulting polaron diameters at room temperature exceed the $MAPbBr_3$ nanocrystal size. This tendency allows the LO-phonon vibrationally excited polaronic quasiparticles to travel through a layer of 3D $MAPbBr_3$ nanocrystals without scattering on the electrostatic potential fluctuations associated with structural imperfections. Accordingly, the mobility and diffusion of polaronic quasiparticles in a layer $MAPbX_3$ nanocrystals at room temperature should be significantly enhanced due to LO-phonon bottleneck.

## IV. CONCLUSIONS

In this article we highlight several basic approaches which would be interesting to a broad audience of scholars exploring unique PL and transport properties of $MAPbX_3$ materials. One of them suggests that two-photon-excited PL spectroscopy and electric-field-induced one-photon-excited PL spectroscopy are required to study the structural phase transitions in 3D $MAPbX_3$ nanocrystals. These techniques are capable of more precisely monitoring the structural phase transitions because the second-order and third-order nonlinear susceptibilities govern the light absorption rates.

Consequently, one can recognize that the structural phase transitions in 3D $MAPbBr_3$ nanocrystals may occur at about the same temperatures as those in single-crystal $MAPbBr_3$. However, the orthorhombic-to-tetragonal structural phase transition in 3D $MAPbBr_3$ nanocrystals, unlike in single-crystal $MAPbX_3$, is spread out over the broad temperature range of $T$ ~70 - 140 K due to the structural phase instability induced by local field fluctuations when free rotations of MA ions are no longer restricted strongly by long-range polar order. The resulting configurational entropy loss and the liquid-like motion of MA cations in 3D $MAPbBr_3$ nanocrystals can be even enhanced by the interfacial electric field arising due to charge separation at the $MAPbBr_3$/ZnO heterointerface, extending the range of the orthorhombic-to-tetragonal structural phase instability from $T$ ~70 to 230 K and significantly shifting the tetragonal-to-cubic phase transition towards higher temperatures exceeding room temperature.

The latter effect is found to be dependent on the ZnO layer thickness and the photoexcited carrier density, thus allowing one to control the structural phase instability range in 3D $MAPbBr_3$ nanocrystals. Finally, we conclude that a stepwise shift of the PL band with temperature observed for single-crystal $MAPbX_3$ is no longer an indication of the structural phase transition in 3D $MAPbBr_3$ nanocrystals because of negligible distortions of $PbX_6$ octahedra under the structural phase instability regime. On the contrary, the nearly monotonic blue shift of PL band with increasing temperature in a fully encapsulated single layer of 20-nm-sized 3D $MAPbBr_3$ nanocrystals seems to result rather from the heating effect under TO/LO phonon bottleneck than that being induced by the progressive $PbX_6$ octahedra distortions.

Furthermore, we point out that two-photon-excited PL spectroscopy and electric-field-induced one-photon-excited PL spectroscopy mainly characterize PL excitation in the nanocrystal core, contrary to conventional one-photon-excited PL spectroscopy dealing with PL excitation in both the nanocrystal core and the nanocrystal surface. Consequently, because the ratio of the surface-to-core PL contributions for nanocrystals is large enough and because the surface states are less sensitive to the structural phase transitions in the core, the structural phase transitions in $MAPbX_3$ nanocrystals can be significantly masked by PL from the surface states when conventional one-photon excitation is applied. This circumstance together with the structural phase instability occurring within the broad temperature range seems to be a reason why the structural phase transitions in $MAPbX_3$ nanocrystals have never been observed.

We also confirmed that the photoexcited carriers responsible for the light-emitting and transport properties of a layer of 3D $MAPbBr_3$ nanocrystals are the polaronic quasiparticles, which can be TO/LO-phonon vibrationally excited to the higher-energy states owing to TO/LO-phonon bottleneck. Consequently, PL from $MAPbBr_3$ nanocrystals results from the recombination of *PE*s, which can emit light either in the ground or TO/LO-phonon vibrationally excited states, thus giving rise to the ~100 meV blue-shift of PL-peak usually appearing in $MAPbBr_3$ nanocrystals with increasing temperature. We note that this polaronic nature of the edge states in $MAPbX_3$ nanocrystals becomes dominant exclusively at higher temperatures (including room temperature) just because energies of the TO/LO-phonon vibrationally excited polaronic quasiparticles (~100 meV) significantly exceeds the ground-state polaron ($\lambda_e$ , $\lambda_h \leq$ ~40 meV) and Rashba energies ($E_R$ ~40 meV). Alternatively, the Rashba spin-split nature of the edge states in $MAPbX_3$ nanocrystals is expected to be dominant only at low temperature when the Rashba energy might exceed energies of the TO/LO-phonon vibrationally excited polaronic quasiparticles.

Additionally, we showed that at room temperature owing to LO-phonon bottleneck, the polaron masses diminish and polaron radii increase. This behavior creates unique conditions for the TO/LO-phonon vibrationally excited polaronic quasiparticles to travel long distances without scattering on electrostatic potential fluctuations governed by structural imperfections.

## ACNOWLEDGEMENTS

This work was supported by the National Key Research and Development Program of China administrated by the Ministry of Science and Technology of China (No. 2016YFB0401702), the Guangdong High-Level Science and Technology Discipline Development Project, Guangdong University Key Laboratory for Advanced Quantum Dot Displays and Lighting (No. 2017KSYS007), the National Natural Science Foundation of China with Grant Nos.11574130 and 61674074, Development and Reform Commission of Shenzhen Project (No. [2017]1395), Shenzhen Peacock Team Project (No. KQTD2016030111203005), Shenzhen Key Laboratory for Advanced quantum dot Displays and Lighting (No. ZDSYS201707281632549), Shenzhen Innovation Project (No. JCYJ20160301113356947, JCYJ20150930160634263, JCYJ20160301113356947). R. Chen acknowledges the support from national 1000 plan for young talents.

**Author contributions**
Y.D.G.: proposed the method of fully encapsulating $MAPbBr_3$ nanocrystals between the two sapphire plates. R.Cai: prepared all the samples, participated in the SEM, TEM, XRD characterization, and in paper writing. Y.D.G. and X.G.: designed the optical experiments and created the optical setup. X.G. and D.W.: performed all the optical experiments. D.W.: performed the SEM, TEM, and XRD characterization experiments, initially treated the optical experiment data, and participated in paper writing. The optical measurements were performed in the laboratory hosted by R.Chen. Y.D.G: mentored students and postdocs, analyzed the data, performed theoretical treatment and numerical simulations and wrote the paper. X.W.S.: guided the research and supervised the project. All authors contributed to discussions.

**Additional information**
**Supplementary Information** accompanies this paper at http://www.nature.com

**Competing financial interests:** The authors declare no competing financial interests.